\documentclass[11pt]{article}
\pdfoutput=1
\usepackage{epsfig,amsmath,amssymb,bm}
\usepackage{graphicx,psfrag}
\usepackage{comment}
\usepackage{textcomp}
\usepackage{hyperref}
\usepackage{xcolor,ulem,braket}
\usepackage[mathscr]{eucal}
\usepackage[top=2cm,bottom=2.5cm,left=2.2cm,right=2.2cm]{geometry}
\usepackage[titletoc]{appendix}
\usepackage[makeroom]{cancel}
\usepackage{enumitem}
\usepackage{stackrel}
\usepackage{float}
\usepackage{mathrsfs}
\usepackage{parskip}
\usepackage[final,color]{showkeys} 
\allowdisplaybreaks

\numberwithin{equation}{section}

\newcommand{\bq}{\begin{equation}}
\newcommand{\eq}{\end{equation}}
\newcommand{\bqa}{\begin{eqnarray}}
\newcommand{\eqa}{\end{eqnarray}}
\newcommand{\nn}{\nonumber \\}

\usepackage{braket}

\def\be{\begin{equation}}
\def\ee{\end{equation}}
\def\bea{\begin{eqnarray}}
\def\eea{\end{eqnarray}}
\def\bnn{\begin{eqnarray*}}
\def\enn{\end{eqnarray*}}

\begin{document}
	
\begin{titlepage}
	\begin{center}
		\hfill 
		\\[22mm]
		{\huge Dual holography as functional renormalization group}\\
		\vspace*{16mm}
		\textbf{Ki-Seok Kim$^{1,2}$, Arpita Mitra$^{3}$, Debangshu Mukherjee$^{1}$ and Seung-Jong Yoo$^{1}$}\\
		\vspace{3mm}
		\begin{center}
			{\it $^{1}$Department of Physics, Pohang University of Science \& Technology (POSTECH), Pohang, Gyeongsangbuk 37673, Korea. \\}
			\vspace*{0.3cm}
			{\it $^{2}$Asia Pacific Center for Theoretical Physics (APCTP), Pohang, Gyeongsangbuk 37673, Korea.\\}
			\vspace*{0.3cm}
			{\it $^{3}$Centre for Theoretical Physics \& Natural Philosophy, Nakhonsawan Studiorum for Advanced Studies, Mahidol University, Thailand.}    
		\end{center}
		\vspace*{1cm}
		\centerline{\footnotesize Email: \texttt{tkfkd@postech.ac.kr}, \texttt{arpita.mit@mahidol.ac.in}, \texttt{debangshum@postech.ac.kr}, \texttt{y2ysj@postech.ac.kr}}
		\date{\today}
	\end{center}
	\medskip
	\vspace{33mm}
		
\begin{abstract}
	We investigate the relationship between the functional renormalization group (RG) and the dual holography framework in the path integral formulation, highlighting how each can be understood as a manifestation of the other. Rather than employing the conventional functional RG formalism, we consider a functional RG equation for the probability distribution function, where the RG flow is governed by a Fokker-Planck-type equation. The central idea is to reformulate the solution of Fokker-Planck type functional RG equation in a path integral representation. Within the semiclassical approximation, this leads to a Hamilton-Jacobi equation for an effective renormalized on-shell action. We then examine our framework for an Einstein-Hilbert action coupled to a scalar field. Applying standard techniques, we derive a corresponding functional RG equation for the distribution function, where the dual holographic path integral serves as its formal solution. By synthesizing these two perspectives, we propose a generalized dual holography framework in which the RG flow is explicitly incorporated into the bulk effective action. This generalization naturally introduces RG $\beta$-functions and reveals that the RG flow of the distribution function is essentially identical to that of the functional RG equation.
\end{abstract}

\end{titlepage}

\newpage 
{\begin{tableofcontents}
\end{tableofcontents}}

\section{Introduction}

Dual holography framework serves as a nonperturbative method in the description of strongly correlated systems. Although string theory gives us the microscopic foundation for the dual holography description \cite{Holographic_Duality_I,Holographic_Duality_II,Holographic_Duality_III,Holographic_Duality_IV}, there have been extensive researches to derive the dual holography framework from quantum field theory (QFT) explicitly. One of the promising direction is based on renormalization group (RG) \cite{SungSik_Holography_I,SungSik_Holography_II,SungSik_Holography_III,SungSik_Holography_IV, Brute_Force_RG_Derivation_Lattice_Kim,Brute_Force_RG_Derivation_Dirac_Kim,Kitaev_Entanglement_Entropy_Kim,RG_GR_Geometry_I_Kim,RG_GR_Geometry_II_Kim,Kondo_Holography,RG_Flow_Direct_Calculation, Nonperturbative_Wilson_RG,Nonperturbative_Wilson_RG_Disorder,Nonperturbative_RG_Flow}. Here, the extra dimension is identified with an RG scale. The so-called holographic renormalization \cite{Holographic_Duality_V,Holographic_Duality_VI,Holographic_Duality_VII,Holographic_Duality_VIII,Holographic_RG_Flow_Ricci_Flow_I,HJ_Review} serves as a general framework to determine the renormalized on-shell effective action, which can be systematically described by the Hamilton-Jacobi equation approach. One may map the Hamilton-Jacobi equation in the bulk into the local RG equation in the boundary \cite{Local_RG_I,Local_RG_II,Local_RG_III} using Hamilton's equation of motion and the IR boundary condition \cite{RG_Monotonicity_NEQ,RG_Flow_Nonperturbative_String}. This is essentially the Callan–Symanzik equation \cite{Peskin_Schroeder_QFT_Textbook} for the renormalized on-shell action of the boundary QFT.

Although a brute-force application of Wilsonian RG transformations have been demonstrated to give the holographic dual description \cite{SungSik_Holography_I,SungSik_Holography_II,SungSik_Holography_III,SungSik_Holography_IV, Brute_Force_RG_Derivation_Lattice_Kim,Brute_Force_RG_Derivation_Dirac_Kim,Kitaev_Entanglement_Entropy_Kim,RG_GR_Geometry_I_Kim,RG_GR_Geometry_II_Kim,Kondo_Holography,RG_Flow_Direct_Calculation, Nonperturbative_Wilson_RG,Nonperturbative_Wilson_RG_Disorder,Nonperturbative_RG_Flow}, an approach based on the so-called \textit{multiscale entanglement renormalization ansatz} (MERA) \cite{MERA} opens a novel direction of research for the dual holography framework \cite{MERA_AdS_CFT}. Inspired by the MERA prescription, as far as we understand, the dual holography framework has been proposed to be the quantum error correction code \cite{QEC_I,QEC_II,QEC_III}. This quantum information perspective serves as an alternate novel understanding of dual holography in addition to the microscopic string theory construction. However, we believe that the connection between the quantum error correction code and the Wilsonian RG framework has not been clarified in these developments. Recently, investigations have shown that the functional RG framework can be viewed as an approximate quantum error correction code \cite{QEC_RG_Intro,QEC_RG_I,QEC_RG_II,QEC_RG_III,QEC_RG_IV}, showing that the Knill-Laflamme condition \cite{KL_Condition} is satisfied at least at the perturbation level of the RG flow.

In this study, we construct the holographic dual description from the functional RG equation \cite{FRG_I,FRG_II,FRG_III}, representing the formal solution of the functional RG equation as a path integral. The path integral reformulation for the functional RG equation gives us a clue on how to generalize the AdS$_{d+1}$/CFT$_{d}$ correspondence, incorporating the information of the RG flow, i.e. RG $\beta-$functions into the bulk effective action of gravity. As a result, we propose a generalized dual holography framework to take the RG flow, consistent with the functional RG equation \cite{RG_Monotonicity_NEQ,RG_Flow_Nonperturbative_String}. 

\section{Path integral reformulation of the functional renormalization group equation}
 
In this section, we introduce a functional RG equation for the probability distribution function and derive a path integral expression as a formal solution. This path integral reformulation will give us a clue on how to incorporate RG flow into the dual holography framework.

\subsection{A review on the functional renormalization group equation}
\label{sec:Review}	

We review the functional RG equation based on ref. \cite{FRG_ML}. The central object of interest is the probability distribution \textit{functional}, schematically given by
\bqa 
&& P_{\Lambda}[\phi(x)] = \frac{1}{Z_{\Lambda}} e^{- S_{\Lambda}[\phi(x)]}\ , \label{Probability_Def}
\eqa
which flows as a function of the momentum cutoff scale $\Lambda$. $\phi(x)$ in Eq. \eqref{Probability_Def} represents a field configuration for a given theory. In this respect, $P_{\Lambda}[\phi(x)]$ may be regarded as the probability density assigned to the field configuration $\phi(x)$ at the scale $\Lambda$. In Eq. (\ref{Probability_Def}),
\bqa
Z_{\Lambda} = \int D \phi(x)\ e^{-S_{\Lambda}[\phi(x)]}
\eqa
is the usual partition function for normalization of the probability density, and $S_{\Lambda}[\phi(x)]$ is an effective action at the scale $\Lambda$. To perform the functional RG analysis, Polchinski wrote down the effective action in the following way	
\bqa && S_{\Lambda}[\phi] = \frac{1}{2} \int \frac{d^{d} p}{(2\pi)^{d}} \phi(p) G^{-1}(p^{2}) K_{\Lambda}^{-1}(p^{2}) \phi(-p) + S^{int}_{\Lambda}[\phi]\ , \label{Effective_Action_Polchinski} 
\eqa
where $\phi(p)$ represents a field configuration in the momentum space. The first term corresponds to a free field theory with a propagator $G(p^{2})$ and a smooth cutoff function $K_{\Lambda}^{-1}(p^{2})$, which suppresses the contribution of momentum modes above the cutoff scale $\Lambda$ by vanishing for $p > \Lambda$. $S^{int}_{\Lambda}[\phi]$ includes various types of interaction vertices that are responsible for the RG flow.

The only guiding principle for the RG flow is the so called unitarity condition,
\bqa && \frac{d }{d \ln \Lambda} \int D \phi\ P_{\Lambda}[\phi] = 0 , \label{Unitarity_Probability} \eqa
i.e., a fixed normalization during the RG flow, which also guarantees that all correlation functions are preserved below the scale $\Lambda$. Inserting Eq. (\ref{Probability_Def}) and Eq. (\ref{Effective_Action_Polchinski}) in Eq. (\ref{Unitarity_Probability}), Polchinski found an exact RG flow equation for $S^{int}_{\Lambda}[\phi]$, where $K_{\Lambda}^{-1}(p^{2})$ has a prescribed dependence on $\Lambda$. Polochinski's equation can be reformulated as a Fokker-Planck type functional differential equation in terms of the probability distribution functional $P_{\Lambda}[\phi]$ \cite{RG_Flow_Relative_Entropy_Gradient_Flow, FRG_ML} as follows	
\begin{equation}
	\begin{aligned}
		& \frac{d }{d \ln \Lambda} P_{\Lambda}[\phi]\\ &= \int_{M} d^{d} x \int_{M} d^{d} y \Big\{ C_{\Lambda}^{Pol.}(x,y) \frac{\delta^{2} P_{\Lambda}[\phi]}{\delta \phi(x) \delta \phi(y)} + \frac{\delta}{\delta \phi(x)} \Big( P_{\Lambda}[\phi] C_{\Lambda}^{Pol.}(x,y) \frac{\delta V_{\Lambda}^{Pol.}[\phi]}{\delta \phi(y)} \Big) \Big\}\\ &\equiv \Delta P_{\Lambda}[\phi] + \mbox{div} \Big( P_{\Lambda}[\phi] \mbox{grad}_{C_{\Lambda}^{Pol.}} V_{\Lambda}^{Pol.}[\phi] \Big)\ . \label{FRG_Eq}   
	\end{aligned}
\end{equation}
where the ERG kernel $ C_{\Lambda}^{Pol.}(x,y)$ plays the role of diffusion constant and $V_{\Lambda}^{Pol.}[\phi(x)]$ is the two-point irreducible vertex which acts as a drift potential. In momentum space, they have the following expression
\begin{equation}\label{CPoldefinition}
	\begin{aligned}
		 C_{\Lambda}^{Pol.}(p^{2}) &= (2\pi)^{d} G(p^{2}) \frac{\partial K_{\Lambda}(p^{2})}{\partial \ln \Lambda}\ ,\\ V_{\Lambda}^{Pol.} &=\int \frac{d^{d} p}{(2\pi)^{d}} \phi(p) G^{-1}(p^{2}) K_{\Lambda}^{-1}(p^{2}) \phi(-p)\ .  
	\end{aligned}
\end{equation}
Furthermore, Eq. \eqref{FRG_Eq} also highlights the Fokker-Planck type structure of the differential equation, where we denote
\bea
\Delta &\equiv& \int_{M} d^{d} x \int_{M} d^{d} y~ C_{\Lambda}^{Pol.}(x,y) \frac{\delta^{2} }{\delta \phi(x) \delta \phi(y)}\ ,\\
\mbox{grad}_{C_{\Lambda}^{Pol.}} &\equiv& \int_{M} d^{d} y~ C_{\Lambda}^{Pol.}(x,y) \frac{\delta }{\delta \phi(y)}\ ,\\
\mbox{div} &\equiv& \int_{M} d^{d} x \frac{\delta}{\delta \phi(x)}\ .
\eea
This Markovian nature of the functional RG equation might be natural, recalling that integrating out high-energy modes erases the memory to renormalize the low-energy dynamics only in the next step of the Wilsonian RG procedure \cite{Brute_Force_RG_Derivation_Lattice_Kim,Brute_Force_RG_Derivation_Dirac_Kim,Kitaev_Entanglement_Entropy_Kim,RG_GR_Geometry_I_Kim,RG_GR_Geometry_II_Kim, Kondo_Holography,RG_Flow_Direct_Calculation,Nonperturbative_Wilson_RG,Nonperturbative_Wilson_RG_Disorder,Nonperturbative_RG_Flow}.

It is straightforward to translate the Fokker-Planck type functional RG equation into a local conservation law,
\bqa && \frac{d }{d \ln \Lambda} P_{\Lambda}[\phi] = - \int_{M} d^{d} x ~\frac{\delta }{\delta \phi(x)} \Big(\Psi_{\Lambda}[\phi;x] P_{\Lambda}[\phi]\Big)\ , \label{Wegner_Morris_Eq} \eqa
where the conserved current is given by
\bqa && \Psi_{\Lambda}[\phi;x] ~P_{\Lambda}[\phi] = - \int_{M} d^{d} y \Big\{ C_{\Lambda}^{Pol.}(x,y) \frac{\delta P_{\Lambda}[\phi]}{\delta \phi(y)}\nonumber\\
&&\hspace*{5cm}+  P_{\Lambda}[\phi]~ C_{\Lambda}^{Pol.}(x,y) \frac{\delta V_{\Lambda}^{Pol.}[\phi]}{\delta \phi(y)}  \Big\}\ . \label{Conserved_Current} \eqa

This local conservation law reproduces the unitarity condition, Eq. (\ref{Unitarity_Probability}), as expected,
\bqa && \frac{d }{d \ln \Lambda} \int D \phi\ P_{\Lambda}[\phi] = - \int D \phi \int_{M} d^{d} x \frac{\delta }{\delta \phi(x)} \Big(\Psi_{\Lambda}[\phi;x] P_{\Lambda}[\phi]\Big)=0\ . \eqa

A characteristic feature of this Fokker-Planck type functional RG equation is that the conserved current is given by the gradient of a functional flow. Therefore $\Psi_{\Lambda}[\phi;x]$ can be represented as a gradient flow of $\Sigma_{\Lambda}[\phi;P_{\Lambda}]$,
\bqa && \Psi_{\Lambda}[\phi;x] = \int_{M} d^{d} y~ C_{\Lambda}(x,y) \frac{\delta \Sigma_{\Lambda}[\phi;P_{\Lambda}]}{\delta \phi(y)} \equiv \mbox{grad}_{C_{\Lambda}^{Pol.}} \Sigma_{\Lambda}[\phi;P_{\Lambda}]\ . \label{Gradient_Flow} \eqa

Reformulation of Eq. (\ref{Conserved_Current}) as the above gradient flow in Eq. (\ref{Gradient_Flow}), one can find an explicit expression of $\Sigma_{\Lambda}[\phi;P_{\Lambda}]$ as
\bqa && \Sigma_{\Lambda}[\phi;P_{\Lambda}] = - \ln \left(\frac{P_{\Lambda}[\phi]}{e^{- V_{\Lambda}^{Pol.}[\phi]}}\right) = S_{\Lambda}[\phi] - V_{\Lambda}^{Pol.}[\phi]\ . \label{KL_Divergence} \eqa 
This $\Sigma_{\Lambda}[\phi;P_{\Lambda}]$ functional turns out to be the Kullback–Leibler (KL) divergence or relative entropy. It plays a central role in the monotonicity of RG flow or in entropy production \cite{FRG_ML,RG_Flow_Relative_Entropy_I,RG_Flow_Relative_Entropy_II,RG_Flow_Relative_Entropy_Gradient_Flow,RG_Flow_Relative_Entropy_0}. Later, we identify the analog of relative entropy in the dual holography framework.

\subsection{Path integral formulation}
\label{sec:functionalRGPI}	

The dual holography framework involves the construction of a dual effective holographic field theory beyond the Wilsonian RG formulation \cite{SungSik_Holography_I,SungSik_Holography_II,SungSik_Holography_III,SungSik_Holography_IV, Brute_Force_RG_Derivation_Lattice_Kim,Brute_Force_RG_Derivation_Dirac_Kim,Kitaev_Entanglement_Entropy_Kim,RG_GR_Geometry_I_Kim,RG_GR_Geometry_II_Kim,Kondo_Holography,RG_Flow_Direct_Calculation, Nonperturbative_Wilson_RG,Nonperturbative_Wilson_RG_Disorder,Nonperturbative_RG_Flow, RG_Flow_Nonperturbative_String}. Given the equivalence between 1-loop RG flow equations in the presence of stochastic noise and the Langevin equation, one can construct a partition function by including the RG flow equations as Faddeev-Popov `gauge' constraints. Furthermore, the $\delta$ function constraints can be averted by introducing Lagrange multiplier fields which consequently act as canonical momentum along the RG direction. Now, the RG scale can be identified with the holographic direction in the emergent bulk and fields in boundary can be upgraded to the fields in the emergent bulk. In this framework, the RG $\beta$-function is given by the gradient of the effective potential originating from integrating out high energy modes.

To verify that the dual holography framework is a path integral reformulation for the functional RG equation, it is necessary to represent a solution of the functional RG equation Eq. (\ref{FRG_Eq}) in the formal path integral expression. Instead of considering the probability density, we focus on the generating functional or partition function for the comparison with the dual holography framework. Although one can derive the generating functional or the partition function of the path integral representation from the Fokker-Planck type functional RG equation directly, here we derive it from the corresponding stochastic (Langevin type) differential equation \cite{MSR_Formulation_SUSY_i,MSR_Formulation_SUSY_ii,MSR_Formulation_SUSY_iii,MSR_Formulation_SUSY_iv,MSR_Formulation_SUSY_v,MSR_Formulation_SUSY_vi,Schwinger_Keldysh_Symmetries_i,Schwinger_Keldysh_Symmetries_ii,Schwinger_Keldysh_Symmetries_iii,Schwinger_Keldysh_Symmetries_iv,Schwinger_Keldysh_Symmetries_v,Schwinger_Keldysh_Symmetries_vi}. Following the standard procedure in the stochastic dynamics, one may consider the following Langevin type differential equation \cite{FRG_ML},
    \bqa && \frac{\partial \phi(x)}{\partial \ln \Lambda} = - \int_{M} d^{d} y~ C_{\Lambda}^{Pol.}(x,y) \frac{\delta V_{\Lambda}^{Pol.}[\phi]}{\delta \phi(y)} + \int_{M} d^{d} y~ \sigma_{\Lambda}(x,y) \frac{\partial \mathcal{W}_{\Lambda}(y)}{\partial \ln \Lambda}\ . \label{Langevin_Eq} \eqa
Here, $\mathcal{W}_{\Lambda}(x)$ is a function valued Wiener process \cite{Wiener_Process}, and $\sigma_{\Lambda}(x,y)$ is the diffusivity kernel defined by the property that it squares to the covariance $C_{\Lambda}^{Pol.}(x,y)$,
\bqa && \int_{M} d^{d} z~ \sigma_{\Lambda}(x,z)~ \sigma_{\Lambda}(z,y) = C_{\Lambda}^{Pol.}(x,y) . \eqa

To clarify the below formal development, one may consider 
\bqa && d \mathcal{W}_{\Lambda}(x) = \xi_{\Lambda}(x) d \ln \Lambda , \eqa
where the white noise correlation is given by	
\bqa \Big\langle \xi_{\Lambda}(x) \xi_{\Lambda'}(y) \Big\rangle = \delta^{(d)}(x-y) \delta(\ln \Lambda-\ln \Lambda')\ .\eqa
Then, we have the RG flow as the Langevin equation,
\bqa && \frac{\partial \phi(x)}{\partial \ln \Lambda} = - \int_{M} d^{d} y~ C_{\Lambda}^{Pol.}(x,y) \frac{\delta V_{\Lambda}^{Pol.}[\phi]}{\delta \phi(y)} + \int_{M} d^{d} y~ \sigma_{\Lambda}(x,y) \xi_{\Lambda}(y)\ .\label{Langevin_Eq_Noise} \eqa

To construct a generating functional associated with Eq. (\ref{Langevin_Eq}) or Eq. (\ref{Langevin_Eq_Noise}), we consider the following $\delta-$function identity,
\bqa 1 &=& \int D \phi\ \delta\left(\frac{\partial \phi(x)}{\partial \ln \Lambda} + \int_{M} d^{d} y\ C_{\Lambda}^{Pol.}(x,y) \frac{\delta V_{\Lambda}^{Pol.}[\phi]}{\delta \phi(y)} - \int_{M} d^{d} y \ \sigma_{\Lambda}(x,y) \xi_{\Lambda}(y) \right) \nn  &&\hspace*{1cm} \times \mbox{det}\left(\int d^dz~\Bigg\{ \delta^{(d)}(x-z)\frac{\partial }{\partial \ln \Lambda}  + \int_{M} d^{d} y~ C_{\Lambda}^{Pol.}(x,y) \frac{\delta^{2} V_{\Lambda}^{Pol.}[\phi]}{\delta \phi(z) \delta \phi(y)} \Bigg\}\right) . \eqa

Here, the $\delta$-function constraint is accompanied by a Jacobian factor, represented by the determinant of the Jacobian matrix. Then, the generating functional can be constructed as follows,
\bqa && \mathcal{Z} = \int D \xi_{\Lambda}\ e^{- \frac{1}{2}\int_{\ln \Lambda_{\rm UV}}^{\ln \Lambda_{\rm IR}} d \ln \Lambda \int_{M} d^{d} z \ \xi_{\Lambda}^{2}(z)} \nn && \times \int D \phi~ \delta\left( \frac{\partial \phi(x)}{\partial \ln \Lambda} + \int_{M} d^{d} y \ C_{\Lambda}^{Pol.}(x,y) \frac{\delta V_{\Lambda}^{Pol.}[\phi]}{\delta \phi(y)} - \int_{M} d^{d} y\ \sigma_{\Lambda}(x,y) \xi_{\Lambda}(y) \right) \nn && \times  ~\mbox{det}~\left( \int_{M} d^{d} z\Bigg\{\delta^{(d)}(x-z) \frac{\partial }{\partial \ln \Lambda} +  \int_{M} d^{d} y~ C_{\Lambda}^{Pol.}(x,y)  \frac{\delta^{2} V_{\Lambda}^{Pol.}[\phi]}{\delta \phi(z) \delta \phi(y)}\Bigg\} \right) . \eqa

The first line performs noise averaging, and the last part introduces the information of the Langevin-type RG flow equation into the generating functional. This construction is called the Fadeev-Popov procedure \cite{Peskin_Schroeder_QFT_Textbook}.

The next step involves identifying a holographic radial direction as the RG scale of the dual QFT with $\ln\Lambda\sim r$ and subsequently evolving the fields as functions of $(x,r)$ in the emergent bulk. Introducing a Lagrange multiplier field $\pi(x,r)$ to impose the $\delta-$function constraint and two fermion ghost fields, $c(x,r)$ and $\bar{c}(x,r)$, to take the Jacobian factor, we represent the above expression in the following way,
\bqa && \mathcal{Z} = \int D \phi(x,r) D \pi(x,r) D c(x,r) D \bar{c}(x,r) D \xi(x,r)\ e^{-S_{\xi}-S_{\phi}}
\eqa
where
\begin{equation}
	\begin{aligned}
		S_{\xi}&= \frac{1}{2}\int_{r_{\rm UV}}^{r_{\rm IR}} d r \int_{M} d^{d} x \ \xi^{2}(x,r) 
	\end{aligned}
\end{equation}
and
\begin{equation}
	\begin{aligned}
		&S_{\phi}= \int_{r_{\rm UV}}^{r_{\rm IR}} d r \int d^{d} x \int d^{d} y\\
		& \Bigg\{ \pi(x,r) \Bigg( \frac{\partial \phi(x,r)}{\partial r} \delta^{(d)}(x-y) + C_{\Lambda}^{Pol.}(x,y,r) \frac{\delta V_{\Lambda}^{Pol.}[\phi]}{\delta \phi(y,r)}-\sigma(x,y,r) \xi(y,r) \Bigg)\\
		& + \int d^{d} z~ \bar{c}(z,r) \Big( \delta^{(d)}(x-z) \delta^{(d)}(z-y) \frac{d }{dr} + C_{\Lambda}^{Pol.}(z,y,r) \frac{\delta^{2} V_{\Lambda}^{Pol.}[\phi]}{\delta \phi(x,r) \delta \phi(y,r)} \Big) c(z,r) \Bigg\}\ .
	\end{aligned}
\end{equation}

In this expression, we notice that $\pi(x,r)$ ($\bar{c}(x,r)$) is the canonical momentum of $\phi(x,r)$ ($c(x,r)$). Finally, we perform the noise averaging to obtain
\begin{equation}
	\begin{aligned}
		\mathcal{Z} &= \int D \phi(x,r) D \pi(x,r) D c(x,r) D \bar{c}(x,r)\\& \exp\Big[ - \int_{r_{\rm UV}}^{r_{\rm IR}} dr \int d^{d} x \int d^{d} y \Bigg\{ \pi(x,r) \Bigg( \partial_r\phi(x,r) \delta^{(d)}(x-y) + C_{\Lambda}^{Pol.}(x,y,r) \frac{\delta V_{\Lambda}^{Pol.}[\phi]}{\delta \phi(y,r)} \Bigg)\\& - \frac{1}{2} \pi(x,r) C_{\Lambda}^{Pol.}(x,y,r) \pi(y,r) \\& + \int d^{d} z ~\bar{c}(z,r)\Bigg( \delta^{(d)}(x-z) \delta^{(d)}(z-y) \frac{d }{d r} + C_{\Lambda}^{Pol.}(z,y,r) \frac{\delta^{2} V_{\Lambda}^{Pol.}[\phi]}{\delta \phi(x,r) \delta \phi(y,r)} \Bigg) c(z,r) \Bigg\} \Bigg]\ . \label{FRG_Dual_Holography}   
	\end{aligned}
\end{equation}

We emphasize that the above path integral expression for the RG flow is purely topological, ensured by $\mathcal{N} = 2$ Becchi-Rouet-Stora-Tyutin (BRST) symmetry \cite{MSR_Formulation_SUSY_i,MSR_Formulation_SUSY_ii,MSR_Formulation_SUSY_iii,MSR_Formulation_SUSY_iv,MSR_Formulation_SUSY_v,MSR_Formulation_SUSY_vi,Schwinger_Keldysh_Symmetries_i,Schwinger_Keldysh_Symmetries_ii,Schwinger_Keldysh_Symmetries_iii,Schwinger_Keldysh_Symmetries_iv,Schwinger_Keldysh_Symmetries_v,Schwinger_Keldysh_Symmetries_vi}. Performing the path integral with respect to all the fields, one finds that the functional integral `localizes' into the Langevin type RG flow equation Eq. (\ref{Langevin_Eq}). The microscopic origin of the $\mathcal{N} = 2$ BRST symmetry lies in unitarity and Kubo–Martin–Schwinger (KMS) symmetry of the path integral formulation. Here, unitarity means that the partition function can be normalized to be $1$ during the RG flow, indicating that the path integral formulation is topological. The KMS symmetry is nothing but the symmetry with respect to `effective' time reversal transformation, here from $\ln \Lambda$ to $\ln \Lambda_{f} - \ln \Lambda$, where $\ln \Lambda$ plays the role of time. $\Lambda_{f}$ is the final cutoff corresponding to the end of the RG transformation. Unitarity gives rise to a set of fermionic supercharges, $Q$ and $\bar{Q}$. These fermionic supercharges do not commute with the KMS symmetry, which requires an additional set of supercharges, $D$ and $\bar{D}$, for the closed algebra. Based on these two sets of supercharges, we can construct Ward identities associated with the RG flow \cite{RG_Monotonicity_NEQ}. In nonequilibrium thermodynamics, such Ward identities have been shown to correspond to generalized fluctuation-dissipation theorems \cite{Jarzynski_i,Jarzynski_ii,Crooks_i,Crooks_ii,Crooks_iii}, being applicable away from equilibrium \cite{MSR_Formulation_SUSY_i,MSR_Formulation_SUSY_ii,MSR_Formulation_SUSY_iii,MSR_Formulation_SUSY_iv,MSR_Formulation_SUSY_v,MSR_Formulation_SUSY_vi,Schwinger_Keldysh_Symmetries_i,Schwinger_Keldysh_Symmetries_ii,Schwinger_Keldysh_Symmetries_iii,Schwinger_Keldysh_Symmetries_iv,Schwinger_Keldysh_Symmetries_v,Schwinger_Keldysh_Symmetries_vi}. It would be interesting to investigate the physical meaning of the Ward identities in the context of the RG flow. In particular, it would be great to verify the $c-$ or $a-$theorem \cite{c_theorem,a_theorem,a_theorem_proof} based on these Ward identities from the $\mathcal{N} = 2$ BRST symmetry, which would generalize the Zamolodchikov's proof in the absence of rotational and translational symmetries. Here, we leave this interesting direction for future research.

\subsection{Hamilton's principal function and Hamilton-Jacobi equation}
\label{sec:HJEquation}

For the comparison with the dual holography framework, we consider the semiclassical limit for the partition function, Eq. (\ref{FRG_Dual_Holography}). Here, we focus on the bosonic sector. Recalling the Legendre transformation in Euclidean flat spacetime, 	
\bqa && \mathcal{L} = \int d^{d} x ~ \pi(x,r) \partial_r \phi(x,r) - \mathcal{H} , \eqa
it is straightforward to read out the Hamiltonian from Eq. (\ref{FRG_Dual_Holography}) as follows
\bqa && \mathcal{H} = \int d^{d} x \int d^{d} y \nn &&\hspace*{0.7cm}\Bigg\{ - \pi(x,r) C_{\Lambda}^{Pol.}(x,y,r) \frac{\delta V_{\Lambda}^{Pol.}[\phi]}{\delta \phi(y,r)} + \frac{1}{2} \pi(x,r) C_{\Lambda}^{Pol.}(x,y,r) \pi(y,r) \Bigg\}\ . \eqa
Then, we find the Hamilton's equation of motion for the RG flow,
\bqa && \dot{\phi}= \frac{\partial \mathcal{H}}{\partial \pi}: ~~~ \dot{\phi}(x,r) = \int d^{d} y~ C_{\Lambda}^{Pol.}(x,y,r) \Big( - \frac{\delta V_{\Lambda}^{Pol.}[\phi]}{\delta \phi(y,r)} + \pi(y,r) \Big) , \label{Canonical_Momentum} \\ && \dot{\pi} = - \frac{\partial \mathcal{H}}{\partial \phi}: ~ \dot{\pi}(x,r)= \int d^{d} y \int d^{d} z~ \pi(z,r)~ C_{\Lambda}^{Pol.}(y,z,r) \frac{\delta^{2} V_{\Lambda}^{Pol.}[\phi]}{\delta \phi(x,r) \delta \phi(y,r)} . \eqa
where $~\dot{}~$ denotes the derivative along the holographic radial direction or equivalently the RG direction and we have used the translational invariance of $C_{\Lambda}^{Pol.}$. It is interesting to observe that the original RG flow equation Eq. (\ref{Langevin_Eq}) has been promoted to be the second order differential equation instead of being the first order after the noise averaging. This structure is in parallel with that of the dual holography framework.

For comparison with the dual holography framework, we invert Eq. (\ref{Canonical_Momentum}) to obtain
\bqa 
&& \pi(x,r) =  \int d^{d} y \Bigg(\frac{\delta V_{\Lambda}^{Pol.}[\phi]}{\delta \phi(y)} +[C_{\Lambda}^{Pol.}]^{-1}(x,y,r)\dot{\phi}(y,r)\Bigg). \label{Canonical_Momentum_Inverted} 
\eqa
This canonical momentum can also be expressed as gradient of the Hamilton's principal function $\mathcal{S}[\phi]$ \cite{HJ_Review},
\bqa && \pi = \frac{\delta \mathcal{S}}{\delta \phi}\ . \label{Principal_Function} \eqa
As a result, we obtain from Eq. \eqref{Canonical_Momentum},
\bqa && \dot{\phi}(x,r) = \int d^{d} y~ C_{\Lambda}^{Pol.}(x,y,r) \frac{\delta }{\delta \phi(y)} \Big( \mathcal{S}[\phi] - V_{\Lambda}^{Pol.}[\phi] \Big)\ , \label{FRG_Gradient_Flow} \eqa
Compared with Eq. (\ref{Gradient_Flow}), the conserved current is proportional to $\dot{\phi}(x,r) $, which confirms that the RG flow is a gradient flow. Furthermore, contrasting with Eq. (\ref{KL_Divergence}), one can realize that the Hamilton's principal function $\mathcal{S}[\phi(x)]$ is proportional to the renormalized action $S_{\Lambda}[\phi(x )]$, and the relative entropy functional $\Sigma_{\Lambda}[\phi(x);P_{\Lambda}]$ is given by $\mathcal{S}[\phi(x)] - V_{\Lambda}^{Pol.}[\phi]$.

Finally, we can discuss the Hamilton-Jacobi equation,
\bqa && \mathcal{H}\Bigg(\phi(x);\frac{\delta \mathcal{S}[\phi]}{\delta \phi(x)}\Bigg) + \frac{\partial \mathcal{S}[\phi]}{\partial r} = 0 . \eqa
More concretely, we find the following expression
\bqa
&&\int d^{d} x\int d^{d} y \Big\{ - \frac{\delta \mathcal{S}}{\delta \phi(x,r)} C_{\Lambda}^{Pol.}(x,y,r) \frac{\delta V_{\Lambda}^{Pol.}[\phi]}{\delta \phi(y,r)} + \frac{1}{2} \frac{\delta \mathcal{S}}{\delta \phi(x,r)} C_{\Lambda}^{Pol.}(x,y,r) \frac{\delta \mathcal{S}}{\delta \phi(y,r)} \Big\} \notag\\&&\hspace*{11.8cm}+ \frac{\partial \mathcal{S}}{\partial r} = 0\ . \label{Hamilton_Jacobi_Eq_FRG} \eqa

As we obtain the Hamilton-Jacobi equation from the Schrodinger equation in the semiclassical limit, we also find it from the Fokker-Planck type functional RG equation Eq. (\ref{FRG_Eq}).

\subsection{Remarks on the path integral reformulation of the functional RG equation}
\label{sec:RemarksPIFRG}

Recalling the path integral construction for the formal solution of the Schrodinger equation, we can write down the probability density at IR as follows:
\bqa && P_{\Lambda}[\phi(x,r_{IR});r_{IR}] = \int D \phi(x,r_{UV}) \exp\Big\{ - \int_{r_{UV}}^{r_{IR}} d r \mathcal{\hat{H}}\Big(\frac{\delta}{\delta \phi(x,r)},\phi(x,r)\Big) \Big\} P_{\Lambda}[\phi(x,r_{UV});r_{UV}] . \nn \eqa
Here, one may easily read out the Hamiltonian kernel for the Wilsonian RG transformation from the functional RG equation Eq.  (\ref{FRG_Eq}) as
\bqa \mathcal{\hat{H}}\Big(\frac{\delta}{\delta \phi(x,r)},\phi(x,r)\Big) &=& \int d^{d} x \ \Bigg( - \int d^{d} y \frac{\delta}{\delta \phi(x,r)} C_{\Lambda}^{Pol.}(x,y,r) \frac{\delta V_{\Lambda}^{Pol.}[\phi]}{\delta \phi(y,r)} \nn &&\hspace*{2.5cm}+ \frac{1}{2} \int d^{d} y \frac{\delta}{\delta \phi(x,r)} C_{\Lambda}^{Pol.}(x,y,r) \frac{\delta}{\delta \phi(y,r)} \Bigg)\ . \eqa

$P_{\Lambda}[\phi(x,r_{UV});r_{UV}] = Z_{\Lambda}^{-1} \exp\Big\{- S_{UV}[\phi(x,r_{UV})] \Big\}$ is the probability density at UV, given the field configuration $\phi(x,r_{UV})$, where the UV action is
\bqa && S_{UV}[\phi(p,r_{UV})] = \frac{1}{2} \int \frac{d^{d} p}{(2\pi)^{d}} \phi(p,r_{UV}) G^{-1}(p^{2}) K_{\Lambda}^{-1}(p^{2},r_{UV}) \phi(-p,r_{UV}) + S^{int}_{UV}[\phi(p,r_{UV})] \nn \eqa
with the UV partition function, \bqa && Z_{\Lambda}[\phi(x,r_{UV});r_{UV}] = \int D \phi(x,r_{UV}) \exp\Big\{- S_{UV}[\phi(x,r_{UV})] \Big\}\ . \eqa

Previously, the Hamiltonian kernel has been constructed in the context of MERA, and applied to the UV probability density explicitly \cite{SungSik_Holography_III,SungSik_Holography_IV}. As a result, the Wilsonian RG transformation has been implemented. An interesting point is that resulting quantum corrections have been reformulated in terms of auxiliary fields, here effective bilocal fields (hopping parameters), which RG-flow in a recursive way. This recursive evolution of the auxiliary field could be formulated as the emergence of an extradimension, identified with an RG scale. Based on this microscopically constructed holographic dual effective field theory, refs. \cite{SungSik_Holography_III,SungSik_Holography_IV} could extract out geometrical information. Considering the saddle-point approximation, semiclassical equations of motion have been derived, supported by both UV and IR boundary conditions and self-consistently determined. After such background configurations of fields had been fixed, Gaussian fluctuations have been considered to give normal modes over the corresponding ground state configuration. It was not difficult to read out the AdS geometry from the linearized equation of motion.

In this study, we do not apply the Hamiltonian kernel to the UV probability density explicitly. Instead, we consider the coherent state representation to translate the above Hamiltonian formulation to the Lagrangian one in the path integral representation as follows:
\bqa && P_{\Lambda}[\phi(x,r_{IR});r_{IR}] = \int D \phi(x,r_{UV}) P_{\Lambda}[\phi(x,r_{UV});r_{UV}] \int D \phi(x,r) D \pi(x,r) \nonumber\\&& \exp\Big[- \int_{r_{UV}}^{r_{IR}} dr \int d^{d} x \Big\{ \int d^{d} y \pi(x,r) \Big( \frac{d \phi(x,r)}{dr} \delta^{(d)}(x-y) + C_{\Lambda}^{Pol.}(x,y,r) \frac{\delta V_{\Lambda}^{Pol.}[\phi]}{\delta \phi(y,r)} \Big) \nn && \hspace*{6.5cm}-\frac{1}{2} \int d^{d} y \pi(x,r) C_{\Lambda}^{Pol.}(x,y,r) \pi(y,r) \Big\} \Big] . \eqa
We believe that this path integral reformulation has to give essentially the same result as refs. \cite{SungSik_Holography_III,SungSik_Holography_IV} in the saddle-point approximation, which reproduces mean-field theory results for critical exponents.

It is straightforward to find the Hamiltonian equation of motion,
\bqa && \dot{\phi}(p,r) = - 2 \frac{\partial \ln K_{\Lambda}(p^{2},r)}{\partial r} \phi(p,r) + G(p^{2}) K_{\Lambda}(p^{2},r) \frac{\partial \ln K_{\Lambda}(p^{2},r)}{\partial r} \pi(p,r) , \\ && \dot{\pi}(p,r) = 2 \frac{\partial \ln K_{\Lambda}(p^{2},r)}{\partial r} \pi(p,r) \ , \eqa
where all symbols have been specified explicitly in the Hamiltonian equation of motion of the previous section.

It is easy to obtain the canonical momentum as
\bqa && \pi(p,r) = f(p) K_{\Lambda}^{2}(p^{2},r) , \eqa
where $f(p)$ is an integral constant. Then, the coordinate field is given by 
\bqa && K_{\Lambda}^{2}(p^{2},r) \phi(p,r) = K_{\Lambda}^{2}(p^{2},r_{UV}) \phi(p,r_{UV}) + \frac{1}{5} G(p^{2}) f(p^{2}) \Big(K_{\Lambda}^{5}(p^{2},r) - K_{\Lambda}^{5}(p^{2},r_{UV})\Big) . \nn \eqa
The UV boundary condition is
\bqa && \pi(p,r_{UV}) = G^{-1}(p^{2}) K_{\Lambda}^{-1}(p^{2},r_{UV}) \phi(p,r_{UV}) + \frac{\partial S^{int}_{\Lambda}[\phi(p,r_{UV})]}{\partial \phi(-p,r_{UV})} . \eqa 
As a result, we determine the integration constant $f(p)$ as
\bqa && f(p) = K_{\Lambda}^{-2}(p^{2},r_{UV})\Big( G^{-1}(p^{2}) K_{\Lambda}^{-1}(p^{2},r_{UV}) \phi(p,r_{UV}) + \frac{\partial S^{int}_{\Lambda}[\phi(p,r_{UV})]}{\partial \phi(-p,r_{UV})} \Big) . \eqa
The IR boundary condition is given by
\bqa && \pi(p,r_{IR}) = 0 . \eqa
As a result, we can determine $\phi(p,r_{UV})$ from the UV mean-field theory equation,
\bqa && G^{-1}(p^{2}) K_{\Lambda}^{-1}(p^{2},r_{UV}) \phi(p,r_{UV}) + \frac{\partial S^{int}_{\Lambda}[\phi(p,r_{UV})]}{\partial \phi(-p,r_{UV})} = 0 . \eqa
This fixes the field configuration as a function of the radial coordinate,
\bqa && \phi(p,r) = K_{\Lambda}^{- 2}(p^{2},r) K_{\Lambda}^{2}(p^{2},r_{UV}) \phi(p,r_{UV}) . \eqa
This mean-field theory result is consistent with that of \cite{SungSik_Holography_III,SungSik_Holography_IV}.

\subsection{Nonperturbative generalization}
\label{sec:NonperturbativePIFRG}

As seen in the previous section, it is disappointing to have just a mean-field theory result in this dual holographic construction. Certainly, we need to improve the RG scheme in the microscopic derivation. In this section, we propose a self-consistent RG scheme, introducing an auxiliary field into the holographic dual effective field theory and performing the RG transformation in a self-consistent way. An idea is to replace the Hamiltonian kernel in
\bqa && P_{\Lambda}[\phi(p,r_{IR}),\Sigma(p,r_{IR});r_{IR}] = \int D \phi(p,r_{UV}) D \Sigma(p,r_{UV}) ~ \mathcal{J}[\Sigma(p,r)] \nn && \exp\Big\{ - \int_{r_{UV}}^{r_{IR}} d r \mathcal{\hat{H}}\Big(\frac{\delta}{\delta \phi(p,r)},\phi(p,r),\frac{\delta}{\delta \Sigma(p,r)},\Sigma(p,r)\Big) \Big\} P_{\Lambda}[\phi(p,r_{UV}),\Sigma(p,r_{UV});r_{UV}] \nn \eqa
with
\bqa && \mathcal{\hat{H}}\Big(\frac{\delta}{\delta \phi(p,r)},\phi(p,r),\frac{\delta}{\delta \Sigma(p,r)},\Sigma(p,r)\Big) \nn && = \int \frac{d^{d} p}{(2\pi)^{d}} \ \Bigg( - \frac{\delta}{\delta \Sigma(p,r)} \Sigma(- p,r) + \frac{u}{2} \frac{\delta}{\delta \Sigma(p,r)} \frac{\delta}{\delta \Sigma(- p,r)} \Bigg)\ \nn && + \int \frac{d^{d} p}{(2\pi)^{d}} \ \Bigg( - \frac{\delta}{\delta \phi(p,r)} \phi(- p,r) + \frac{1}{2} \int \frac{d^{d} p'}{(2\pi)^{d}} \ \frac{\delta}{\delta \phi(p,r)} \mathcal{G}[p^{2};\Sigma(p-p',r)] \frac{\delta}{\delta \phi(- p',r)} \Bigg)\ , \nn \label{H} \eqa
where a bilocal field $\Sigma(p,r)$ has been taken into account in the Wilsonian RG flow. The bilocal field $\Sigma(p,r)$ may be regarded as a dual field of $\phi(p,r) \phi(- p,r)$. In the first parentheses on the r.h.s of Eq. \eqref{H}, $u$ is the self-interaction parameter in the $\phi^{4}$ theory. $\mathcal{G}[p^{2};\Sigma(p-p',r)]$ is the Green's function of $\phi(p,r)$, given by	
\bqa && \Big( p^{2} + m^{2} - i \Sigma(p-p',r) \Big) \mathcal{G}[p^{2};\Sigma(p-p',r)] = 1 , \eqa
where the bilocal field has been introduced (See Appendix \ref{appendix1} for a detailed derivation). One can understand the bilocal field as the collective degree of freedom describing 2-point functions and holography emerges only when these correlations become dynamical fields in the extra dimension.

$\mathcal{J}[\Sigma(p,r)]$ is the Jacobian factor, given by
\bqa && \mathcal{J}[\Sigma(p,r)] = \exp\Big\{ - \frac{1}{2} \int_{r_{UV}}^{r_{IR}} d r \int \frac{d^{d} p}{(2\pi)^{d}} \int \frac{d^{d} p'}{(2\pi)^{d}} \ln \mathcal{G}^{-1}[p^{2};\Sigma(p-p',r)] \Big\} . \eqa
Now, the UV probability density $$P_{\Lambda}[\phi(p,r_{UV}),\Sigma(p,r_{UV});r_{UV}] = Z_{\Lambda}^{-1} \exp\Big\{- S_{UV}[\phi(p,r_{UV}),\Sigma(p,r_{UV})] \Big\}$$ also depends on the bilocal field, where the UV action is given by
\bqa S_{UV}[\phi(p,r_{UV}),\Sigma(p,r_{UV})] &=& \int \frac{d^{d} p}{(2\pi)^{d}} \Big\{ \frac{1}{2} \int \frac{d^{d} p'}{(2\pi)^{d}} \phi(p,r_{UV}) \mathcal{G}^{-1}[p^{2};\Sigma(p-p',r_{UV})] \phi(-p',r_{UV}) \nn &&\hspace*{4.5cm}+\frac{1}{2 u} \Sigma(p,r_{UV}) \Sigma(- p,r_{UV}) \Big\}\ . \eqa
with the UV partition function
\begin{equation}
	Z_{\Lambda}[\phi(x,r_{UV});r_{UV}] = \int D \phi(p,r_{UV}) D \Sigma(p,r_{UV}) \exp\Big\{- S_{UV}[\phi(p,r_{UV}),\Sigma(p,r_{UV})] \Big\}\ .
\end{equation}
\newpage
Based on these building blocks, we obtain the following path integral representation for the probability density at IR, 
\bqa && P_{\Lambda}[\phi(p,r_{IR}),\Sigma(p,r_{IR});r_{IR}] = \int D \phi(p,r_{UV}) D \Sigma(p,r_{UV}) P_{\Lambda}[\phi(p,r_{UV}),\Sigma(p,r_{UV});r_{UV}] \nn && \int D \Pi_{\phi}(p,r) D \phi(p,r) D \Pi_{\Sigma}(p,r) D \Sigma(p,r) \nn && \exp\Big[- \int_{r_{UV}}^{r_{IR}} dr \int \frac{d^{d} p}{(2\pi)^{d}} \Big\{ \Pi_{\phi}(p,r) \partial_{r} \phi(p,r) - \frac{1}{2} \int \frac{d^{d} p'}{(2\pi)^{d}} \Pi_{\phi}(p,r) \mathcal{G}[p^{2};\Sigma(p-p',r)] \Pi_{\phi}(-p',r)  \nn && + \Pi_{\Sigma}(p,r) \partial_{r} \Sigma(p,r) - \frac{u}{2} \Pi_{\Sigma}(p,r) \Pi_{\Sigma}(- p,r) + \frac{1}{2} \int \frac{d^{d} p'}{(2\pi)^{d}} \ln \mathcal{G}^{-1}[p^{2};\Sigma(p-p',r)] \Big\} \Big] . \eqa
Focusing on the partition function, and performing $\int D \Pi_{\phi}(p,r) D \phi(p,r) D \Pi_{\Sigma}(p,r)$, we obtain
\bqa && Z = \int D \Sigma(p,r) \exp\left[ - \int \frac{d^{d} p}{(2\pi)^{d}} \left\{ \frac{1}{2} \int \frac{d^{d} p'}{(2\pi)^{d}} \ln \mathcal{G}^{-1}[p^{2};\Sigma(p-p',r_{IR})]\right.\right.\nn
&&\hspace*{9cm}\left. + \frac{1}{2 u} \Sigma(p,r_{UV}) \Sigma(- p,r_{UV}) \right\} \nn && \hspace*{-0.7cm}\left.- \int_{r_{UV}}^{r_{IR}} dr \int \frac{d^{d} p}{(2\pi)^{d}} \Big\{ \frac{1}{2 u} \Big( \partial_{r} \Sigma(p,r) \Big) \Big( \partial_{r} \Sigma(- p,r) \Big) + \frac{1}{2} \int \frac{d^{d} p'}{(2\pi)^{d}} \ln \mathcal{G}^{-1}[p^{2};\Sigma(p-p',r)] \right] . \label{HDEFT_Nonperturbative} \eqa
Here, $\frac{1}{2} \int \frac{d^{d} p'}{(2\pi)^{d}} \ln \mathcal{G}^{-1}[p^{2};\Sigma(p-p',r)]$ originates from $\int D \Pi_{\phi}(p,r) D \phi(p,r)$ while\\
$\frac{1}{2 u} \Big( \partial_{r} \Sigma(p,r) \Big) \Big( \partial_{r} \Sigma(- p,r) \Big)$ results from $\int D \Pi_{\Sigma}(p,r)$. The saddle-point equation is given by the second-order differential equation,
\bqa && \frac{1}{u} \partial_{r}^{2} \Sigma(p,r) = - \frac{i}{2} \int \frac{d^{d} p'}{(2\pi)^{d}} \mathcal{G}[p^{2};\Sigma(p-p',r)] ,\label{eq:dynamical} \eqa
supported by both the IR and UV boundary conditions, respectively,
\bqa && \frac{1}{u} \partial_{r} \Sigma(p,r) \Big|_{r_{IR}} = \frac{i}{2} \int \frac{d^{d} p'}{(2\pi)^{d}} \mathcal{G}[p^{2};\Sigma(p-p',r_{IR})] , \\ && \partial_{r} \Sigma(p,r) \Big|_{r_{UV}} = \Sigma(p,r_{UV}) . \eqa
In particular, Eq. \eqref{eq:dynamical} represents second order flow equation of propagating bulk fields. In appendix, we derive this path integral construction in a brute force way of the Wilsonian RG transformation. We show that the microscopically derived holographic dual effective field theory, Eq. (\ref{HDEFT_Nonperturbative_Local_Approx}) is identical to Eq. (\ref{HDEFT_Nonperturbative}). In this study, we do not discuss possible solutions further.

\section{From the path integral formulation to the Fokker-Planck type functional RG equation in AdS$_{d+1}$/CFT$_{d}$}
\label{sec:gravityRG}

In this section, we perform the reverse engineering from the path integral formulation to the Fokker-Planck type functional RG equation in the AdS$_{d+1}$/CFT$_{d}$ correspondence. In particular, we review the AdS$_{d+1}$/CFT$_{d}$ correspondence in the Hamiltonian formulation, and clarify missing ingredients, compared to the functional RG equation of the previous section.

We consider the Einstein-Hilbert action minimally coupled with a scalar field for concreteness,
\begin{align}
	S = - \frac{1}{2 \kappa^{2}} \Big\{ \int_{\mathcal{M}} d^{d+1} x \sqrt{g} \Big( R[g] - \frac{1}{2} g^{\mu\nu} \partial_{\mu} \phi \partial_{\nu} \phi - m^{2} \phi^{2} - U(\phi) \Big)+ \int_{\partial \mathcal{M}} d^{d} x \sqrt{\gamma} 2 K \Big\} , \label{Einstein_Hilbert_Action}
\end{align}
where the gravitational coupling is given by $\kappa^{2} = 8 \pi G_{d+1}$. $\mathcal{M}$ denotes a bulk manifold with a boundary $\partial \mathcal{M}$. In addition to the Einstein-Hilbert action, we consider the dynamics of a scalar field, where $U(\phi)$ is an effective potential, including the contribution from cosmological constant. Here, we do not take into account the curvature induced mass term, $R[g] \phi^{2}$, for our simple presentation of the Hamiltonian formulation. The last term represents the Gibbons–Hawking–York (GHY) boundary term \cite{GHY_Boundary_Term}, where $K$ is the extrinsic curvature of the boundary.

To figure out how the functional RG framework can be encoded into this effective gravity action, we consider the Hamiltonian formulation for Eq. (\ref{Einstein_Hilbert_Action}), referred to as Arnowitt, Deser, and Misner (ADM) formalism \cite{ADM_Formulation}. Here, we review it based on refs. \cite{HJ_Review,GR_Textbook}. The metric is decomposed as follows
\bqa && d s^{2} = g_{\mu\nu} d x^{\mu} d x^{\nu} = (N^{2} + N_{i} N^{i}) d r^{2} + 2 N_{i} d r d x^{i} + \gamma_{ij} d x^{i} d x^{j} . \eqa
$N$ is the lapse function encoding the RG evolution between ADM RG hypersurfaces, and $N_{i}$ is the shift vector describing how spatial coordinates change between such hypersurfaces. $\gamma_{ij}$ with $i, j = 1, ..., d$ is an induced metric on each hypersurface.

The extrinsic curvature is given by the Lie derivative of the metric along $n^{\mu} = (1/N, - N^{i}/N)$ as follows
\bqa && K_{ij} = \frac{1}{2} (\mathcal{L}_{n} g)_{ij} = \frac{1}{2 N} \Big( \dot{\gamma}_{ij} - D_{i} N_{j} - D_{j} N_{i}\Big) , \eqa
where $D_{i}$ is the covariant derivative on the ADM RG hypersurface and $\dot{\gamma}_{ij} \equiv \partial_{r} {\gamma}_{ij}$. Then, the scalar curvature can be decomposed as 	
\bqa && R[g] = R[\gamma] + K^{2} - K_{ij} K^{ij} + \nabla_{\mu} \zeta^{\mu} , \eqa
where $K = \gamma^{ij} K_{ij}$ and $\zeta^{\mu} = - 2 K n^{\mu} + 2 n^{\rho} \nabla_{\rho} n^{\mu}$. As a result, Eq. (\ref{Einstein_Hilbert_Action}) is expressed in the following Lagrangian	
\bqa && L = - \frac{1}{2 \kappa^{2}} \int_{\Sigma_{r}} d^{d} x \sqrt{\gamma} N \Big\{ R[\gamma] + K^{2} - K_{ij} K^{ij} - \frac{1}{2 N^{2}} \dot{\phi}^{2} + \frac{N^{i}}{N^{2}} \dot{\phi} \partial_{i} \phi \notag\\&&\qquad\qquad- \frac{1}{2} \Big(\gamma^{ij} + \frac{N^{i}N^{j}}{N^{2}}\Big) \partial_{i} \phi \partial_{j} \phi - m^{2} \phi^{2} - U(\phi) \Big\} , \label{ADM_Lagrangian} \eqa where $\Sigma_{r}$ is the ADM RG hypersurface, and the original action is given by $S = \int d r L$.

To construct the Hamiltonian, we introduce the canonical momenta as follows
\bqa && \pi^{ij} = \frac{\partial L}{\partial \dot{\gamma}_{ij}} = - \frac{1}{2 \kappa^{2}} \sqrt{\gamma} (K \gamma^{ij} - K^{ij}) , \label{Canonical_Momentum_Metric} \\ && \pi_{\phi} = \frac{\partial L}{\partial \dot{\phi}} = \frac{1}{2\kappa^{2} N} \sqrt{\gamma} (\dot{\phi} - N^{i} \partial_{i} \phi) . \label{Canonical_Momentum_Scalar} \eqa
$\pi^{ij}$ ($\pi_{\phi}$) is the canonical momentum of the induced metric (scalar field). The other two canonical momenta are given by
\bqa && \pi_{N} = \frac{\partial L}{\partial \dot{N}} = 0\ , ~~~~~ \pi_{N^{i}} = \frac{\partial L}{\partial \dot{N}^{i}} = 0 , \eqa
which correspond to the Hamiltonian constraint and the momentum constraint, respectively.

We perform the Legendre transformation to obtain the Hamiltonian from the Lagrangian as follows
\bqa && H = \int_{\Sigma_{r}} d^{d} x (\pi^{ij} \dot{\gamma}_{ij} + \pi_{\phi} \dot{\phi}) - L = \int_{\Sigma_{r}} d^{d} x (N \mathcal{H} + N_{i} \mathcal{H}^{i}) . \label{ADM_Hamiltonian} \eqa
As a result, we find the following Hamiltonian density and momentum density, respectively,
\begin{align}
	&\mathcal{H} = \frac{2 \kappa^{2}}{\sqrt{\gamma}} \Big( \pi_{ij} \pi^{ij} - \frac{1}{d-1} \pi^{2} + \frac{1}{2} \pi_{\phi}^{2} \Big) + \frac{\sqrt{\gamma}}{2 \kappa^{2}} \Big(R[\gamma] - \frac{1}{2} \gamma^{ij} \partial_{i} \phi \partial_{j} \phi - m^{2} \phi^{2} - U(\phi) \Big) , \\&\mathcal{H}^{i} = - 2 D_{j} \pi^{ij} + \pi_{\phi} \partial^{i} \phi .
\end{align}

To clarify the connection with the RG flow, we consider the gauge fixing condition,
\bqa && N = 1 , ~~~~~ N^{i} = 0 . \eqa
Then, Eq. (\ref{ADM_Hamiltonian}) reads
\bqa && H = \int_{\Sigma_{r}} d^{d} x \mathcal{H} . \eqa
As a result, by plugging Eq. \eqref{ADM_Hamiltonian},  Eq. (\ref{ADM_Lagrangian}) can be expressed as
\bqa && S = \int d r \int_{\Sigma_{r}} d^{d} x \Big\{ \pi^{ij} \dot{\gamma}_{ij} + \pi_{\phi} \dot{\phi} - \frac{2 \kappa^{2}}{\sqrt{\gamma}} \pi^{ij} \Big(\gamma_{ik} \gamma_{jl} - \frac{1}{d-1} \gamma_{ij} \gamma_{kl}\Big) \pi^{kl} - \frac{\kappa^{2}}{\sqrt{\gamma}} \pi_{\phi}^{2}\nn && \hspace*{10cm}- \mathcal{V}_{eff}[\gamma_{ij},\phi] \Big\} , \label{ADM_Action} \eqa 

where the effective potential is given by
\bqa && \mathcal{V}_{eff}[\gamma_{ij},\phi] = \frac{\sqrt{\gamma}}{2 \kappa^{2}} \Big(R[\gamma] - \frac{1}{2} \gamma^{ij} \partial_{i} \phi \partial_{j} \phi - m^{2} \phi^{2} - U(\phi) \Big) . \label{Effective_Potential_AdS_CFT} \eqa

To compare the dual holography framework with the Hamilton-Jacobi equation of the functional RG flow in the previous section, we introduce the Hamilton's principal function $\mathcal{S}[\gamma_{ij},\phi]$. Then, both canonical momenta are given by
\bqa && \pi^{ij} = \frac{\delta \mathcal{S}[\gamma_{ij},\phi]}{\delta \gamma_{ij}} , ~~~~~ \pi_{\phi} = \frac{\delta \mathcal{S}[\gamma_{ij},\phi]}{\delta \phi} . \label{Hamilton_Principal_FT_AdS_CFT} \eqa
As a result, we obtain the Hamilton-Jacobi equation 
\bqa && H\Big(\gamma_{ij}, \phi; \frac{\delta \mathcal{S}[\gamma_{ij},\phi]}{\delta \gamma_{ij}},\frac{\delta \mathcal{S}[\gamma_{ij},\phi]}{\delta \phi}\Big) + \frac{\partial \mathcal{S}[\gamma_{ij},\phi]}{\partial r} = 0\ ,\eqa
which when explicitly written out, takes the form \cite{Holographic_Duality_V,Holographic_Duality_VI,Holographic_Duality_VII,Holographic_Duality_VIII,Holographic_RG_Flow_Ricci_Flow_I,HJ_Review}
\bqa && \int_{\Sigma_{r}} d^{d} x \Bigg\{ \frac{2 \kappa^{2}}{\sqrt{\gamma}} \Big(\gamma_{ik} \gamma_{jl} - \frac{1}{d-1} \gamma_{ij} \gamma_{kl}\Big)  \Big(\frac{\delta \mathcal{S}[\gamma_{ij},\phi]}{\delta \gamma_{ij}}\Big) \Big(\frac{\delta \mathcal{S}[\gamma_{ij},\phi]}{\delta \gamma_{kl}}\Big) + \frac{\kappa^{2}}{ \sqrt{\gamma}} \Big(\frac{\delta \mathcal{S}[\gamma_{ij},\phi]}{\delta \phi}\Big)^{2} \nn && \hspace*{7cm} +\mathcal{V}_{eff}[\gamma_{ij},\phi] \Bigg\}+ \frac{\partial \mathcal{S}[\gamma_{ij},\phi]}{\partial r} = 0\ . \label{Hamilton_Jacobi_Eq_AdS_CFT} \eqa
Here, we note that $\frac{\partial \mathcal{S}[\gamma_{ij},\phi]}{\partial r} = 0$, is a consequence of the Hamiltonian constraint.

Comparing the above with Eq. (\ref{Hamilton_Jacobi_Eq_FRG}),
\bqa && \int d^{d} x \int d^{d} y \Big\{ - \frac{\delta \mathcal{S}}{\delta \phi(x,r)} C_{\Lambda}^{Pol.}(x,y,r) \frac{\delta V_{\Lambda}^{Pol.}[\phi(x,r)]}{\delta \phi(y,r)} + \frac{1}{2} \frac{\delta \mathcal{S}}{\delta \phi(x,r)} C_{\Lambda}^{Pol.}(x,y) \frac{\delta \mathcal{S}}{\delta \phi(y,r)} \Big\} \nonumber\\&&\hspace*{11.8cm}+ \frac{\partial \mathcal{S}}{\partial r} = 0\ , \nonumber \eqa
we observe that there does not exist a term in Eq. (\ref{Hamilton_Jacobi_Eq_AdS_CFT}), which corresponds to\\ $- \frac{\delta \mathcal{S}}{\delta \phi(x)} C_{\Lambda}^{Pol.}(x,y) \frac{\delta V_{\Lambda}^{Pol.}[\phi(x)]}{\delta \phi(y)}$ in Eq. (\ref{Hamilton_Jacobi_Eq_FRG}). This incompatibility leads us to generalize the AdS/CFT correspondence framework as follows: we introduce the RG $\beta-$function into the bulk gravity action, describing the RG flow.

The above incompatibility can be also discussed in the level of the functional RG equation. Following the standard procedure from the Schr\"odinger equation to the Hamilton-Jacobi equation or the prescription in our previous studies \cite{RG_Monotonicity_NEQ,RG_Flow_Nonperturbative_String}, we obtain the functional RG equation, which has the generating functional (partition function) given by the effective action Eq. (\ref{ADM_Action}) as a formal solution, as follows
\bqa && \Big\{ \frac{\partial }{\partial r} - \int_{\Sigma_{r}} d^{d} x \frac{\sqrt{\gamma}}{2 \kappa^{2}} \Big(R[\gamma] - \frac{1}{2} \gamma^{ij} \partial_{i} \phi \partial_{j} \phi - m^{2} \phi^{2} - U(\phi)\Big) \Big\} P_{r}[\gamma_{ij},\phi] \nn && \hspace*{1.1cm}= \int_{\Sigma_{r}} d^{d} x \Big\{ \frac{2 \kappa^{2}}{\sqrt{\gamma}} \frac{\delta }{\delta \gamma_{ij}} \Big(\gamma_{ik} \gamma_{jl} - \frac{1}{d-1} \gamma_{ij} \gamma_{kl}\Big) \frac{\delta }{\delta \gamma_{kl}} + \frac{\kappa^{2}}{\sqrt{\gamma}} \frac{\delta^{2} }{\delta \phi^{2}} \Big\} P_{r}[\gamma_{ij},\phi]\ . \label{FRG_Eq_AdS_CFT} 
\eqa
\\In other words, we find $P_{r}[\gamma_{ij},\phi] \propto e^{- S[\gamma_{ij},\phi]}$, where the effective action is given by Eq. (\ref{ADM_Action}). Therefore, the Hamilton-Jacobi equation in Eq. \eqref{Hamilton_Jacobi_Eq_AdS_CFT} can be regarded as the WKB
approximation to the Schr\"odinger equation in Eq. \eqref{FRG_Eq_AdS_CFT}, where one writes the wave-function as $P_{r}[\gamma_{ij},\phi]$
and keeps only the leading order terms in recalling $\kappa^2\sim G_{d+1}\sim 1/N^2$.

Compared to Eq. (\ref{FRG_Eq}),		
\bqa \begin{aligned}
	\frac{d }{d \ln \Lambda} P_{\Lambda}[\phi] &= \int_{M} d^{d} x \int_{M} d^{d} y \Big\{ C_{\Lambda}^{Pol.}(x,y) \frac{\delta^{2} P_{\Lambda}[\phi]}{\delta \phi(x) \delta \phi(y)}\\
	&\hspace*{5cm}+ \frac{\delta}{\delta \phi(x)} \Big( P_{\Lambda}[\phi] C_{\Lambda}^{Pol.}(x,y) \frac{\delta V_{\Lambda}^{Pol.}[\phi]}{\delta \phi(y)} \Big) \Big\}\ , \nonumber
\end{aligned}
\eqa
we find that there does not exist the RG flow information in Eq. (\ref{FRG_Eq_AdS_CFT}). However, we emphasize that both functional RG flow equations are Markovian, given by the Fokker-Planck type equation.

Before closing this section, we justify the Weyl anomaly interpretation for the effective potential $\mathcal{V}_{eff}[\gamma_{ij},\phi]$ in Eq. (\ref{Hamilton_Jacobi_Eq_AdS_CFT}). Even if the UV theory is invariant under Weyl transformation, the RG flow of the theory induce an explicit breaking of Weyl invariance at intermediate scales giving rise to the Weyl anomaly. The central idea is to recast the Hamilton-Jacobi equation Eq. (\ref{Hamilton_Jacobi_Eq_AdS_CFT}) as the local RG equation in the following way
\bqa && \Bigg\{ \frac{\partial }{\partial r} + \frac{1}{2} \int_{\Sigma_{r}} d^{d} x~ \left(\beta_{ij}\frac{\delta }{\delta \gamma_{ij}} + \beta_{\phi}\frac{\delta }{\delta \phi}\right)\Bigg\} \mathcal{S} = \int_{\Sigma_{r}} d^{d} x~ \mathcal{A}\ . 
\eqa
Here, both RG $\beta-$functions are given by
\bqa && \dot{\gamma}_{ij} = \frac{4 \kappa^{2}}{\sqrt{\gamma}} \Big(\gamma_{ik} \gamma_{jl} - \frac{1}{d-1} \gamma_{ij} \gamma_{kl}\Big) \Big(\frac{\delta \mathcal{S}}{\delta \gamma_{kl}}\Big) \equiv \beta_{ij} , \\ && \dot{\phi} = \frac{2\kappa^{2}}{\sqrt{\gamma}} \frac{\delta \mathcal{S}}{\delta \phi} \equiv \beta_{\phi} , \label{RG_Flow_Scalar_AdS_CFT} \eqa 
which result from the canonical momenta, Eqs. (\ref{Canonical_Momentum_Metric}) and (\ref{Canonical_Momentum_Scalar}) with the Hamilton's principal function of Eq. (\ref{Hamilton_Principal_FT_AdS_CFT}), respectively. On the RHS, $\mathcal{A}$ represents the Weyl anomaly, given by the effective potential $\mathcal{V}_{eff}[\gamma_{ij},\phi]$,
\bqa && \mathcal{A} = \frac{\sqrt{\gamma}}{2 \kappa^{2}} \Big(R[\gamma] - \frac{1}{2} \gamma^{ij} \partial_{i} \phi \partial_{j} \phi - m^{2} \phi^{2} - U(\phi)\Big) . \eqa
If we compare Eq. (\ref{RG_Flow_Scalar_AdS_CFT}) with Eq. (\ref{FRG_Gradient_Flow}), 	
\bqa
&& \frac{d \phi(x)}{d \ln \Lambda} = \int d^{d} y\ C_{\Lambda}^{Pol.}(x,y) \frac{\delta }{\delta \phi(y)} \Big( \mathcal{S}[\phi] - V_{\Lambda}^{Pol.}[\phi] \Big) , \nonumber
\eqa
we find that there is no effective potential term in Eq. (\ref{RG_Flow_Scalar_AdS_CFT}).

\section{Incorporating information of RG flow into the dual holography framework}
\label{sec:incorporatingRG}

In this section, we generalize dual holography to take into account the information of the RG flow at the level of the bulk effective action. As a result, the generalized dual holography framework allows the RG flow description in a nonperturbative way, being consistent with the functional RG equation description of the previous section.

Based on the previous discussion, we introduce the information of the RG flow as follows	
\bqa && \beta_{ij} = \frac{1}{\sqrt{\gamma}} \frac{\partial \mathcal{V}_{eff}[\gamma_{ij},\phi]}{\partial \gamma^{ij}} , ~~~~~ \beta_{\phi} = \frac{1}{\sqrt{\gamma}} \frac{\partial \mathcal{V}_{eff}[\gamma_{ij},\phi]}{\partial \phi} , \eqa
where all these RG $\beta-$functions are given by the gradient of the effective potential, Eq. (\ref{Effective_Potential_AdS_CFT}). Then, we generalize the effective bulk action Eq. (\ref{ADM_Action}) for the AdS$_{d+1}$/CFT$_{d}$ correspondence taking into account the gradient RG flow $\beta-$functions by defining the following effective action	
\bqa && S = \int_{0}^{R} d r \int_{\Sigma_{r}} d^{d} x \Big\{ \pi^{ij} \Big( \dot{\gamma}_{ij} - \beta_{ij} \Big) + \pi_{\phi} \Big( \dot{\phi} - \beta_{\phi} \Big) - \frac{2 \kappa^{2}}{\sqrt{\gamma}} \pi^{ij} \Big(\gamma_{ik} \gamma_{jl} - \frac{1}{d-1} \gamma_{ij} \gamma_{kl}\Big) \pi^{kl}\nn &&\hspace*{5.6cm} - \frac{\kappa^{2}}{\sqrt{\gamma}} \pi_{\phi}^{2} - \mathcal{V}_{eff}[\gamma_{ij},\phi] \Big\} - k_1\int_{\Sigma_{R}} d^{d} x\ \mathcal{V}_{eff}[\gamma_{ij},\phi]\Big|_R . \label{Dual_Holography_General_Framework} \eqa

Here, $r=R$ is the radial slice signifying the end point (i.e. the IR fixed point) of the RG transformation. Besides the introduction of the RG $\beta-$functions, we also considered the effective potential $\int_{\Sigma_{R}} d^{d} x\ \mathcal{V}_{eff}[\gamma_{ij},\phi]$ at the ADM RG hypersurface $\Sigma_{R}$. Actually, this introduction is based on our previous study, where the integration of high energy modes at the RG scale $R$ gives rise to the effective potential at the RG hypersurface (boundary) \cite{Brute_Force_RG_Derivation_Lattice_Kim,Brute_Force_RG_Derivation_Dirac_Kim,Kitaev_Entanglement_Entropy_Kim,RG_GR_Geometry_I_Kim,RG_GR_Geometry_II_Kim,Kondo_Holography,RG_Flow_Direct_Calculation,
	Nonperturbative_Wilson_RG,Nonperturbative_Wilson_RG_Disorder,Nonperturbative_RG_Flow}. The effective boundary potential does not affect anything in the bulk equations of motion but changes the IR boundary condition. We will see that this introduction is consistent with the functional RG equation, more precisely, the gradient RG flow.
%
%
Accordingly, the quantum partition function is given by
\begin{align}\label{eq:RGaction}
	Z &= \int D \gamma_{ij} D \pi^{ij} D \phi D \pi_{\phi} \exp\Big[ - \int_{0}^{R} d r \int_{\Sigma_{r}} d^{d} x \Big\{ \pi^{ij} \Big( \dot{\gamma}_{ij} - \beta_{ij} \Big) + \pi_{\phi} \Big( \dot{\phi} - \beta_{\phi} \Big) \notag\\& - \frac{2 \kappa^{2}}{\sqrt{\gamma}} \pi^{ij} \Big(\gamma_{ik} \gamma_{jl} - \frac{1}{d-1} \gamma_{ij} \gamma_{kl}\Big) \pi^{kl} - \frac{\kappa^{2}}{\sqrt{\gamma}} \pi_{\phi}^{2} - \mathcal{V}_{eff}[\gamma_{ij},\phi] \Big\} +k_1 \int_{\Sigma_{R}} d^{d} x\ \mathcal{V}_{eff}[\gamma_{ij},\phi]\Big|_R \Big] . \end{align}

%
%

It is straightforward to find the corresponding Hamilton-Jacobi equation. Considering Hamilton's principal function for the canonical momenta as
\bqa && \pi^{ij} = \frac{\delta \mathcal{S}[\gamma_{ij},\phi]}{\delta \gamma_{ij}}\ , ~~~~~ \pi_{\phi} = \frac{\delta \mathcal{S}[\gamma_{ij},\phi]}{\delta \phi}\ , \eqa
we obtain the Hamilton-Jacobi equation,
\bqa && H\Big(\gamma_{ij}, \phi; \frac{\delta \mathcal{S}[\gamma_{ij},\phi]}{\delta \gamma_{ij}},\frac{\delta \mathcal{S}[\gamma_{ij},\phi]}{\delta \phi}\Big) + \frac{\partial \mathcal{S}[\gamma_{ij},\phi]}{\partial r} = 0 . \eqa
Here, the Hamiltonian density is modified as
\begin{align}
	\mathcal{H} = \frac{2 \kappa^{2}}{\sqrt{\gamma}} \pi^{ij} \Big(\gamma_{ik} \gamma_{jl} - \frac{1}{d-1} \gamma_{ij} \gamma_{kl}\Big) \pi^{kl} + \pi^{ij} \beta_{ij} + \frac{\kappa^{2}}{ \sqrt{\gamma}} \pi_{\phi}^{2} + \pi_{\phi} \beta_{\phi} + \mathcal{V}_{eff}[\gamma_{ij},\phi]\ , 
\end{align}
where contributions from $\pi^{ij} \beta_{ij}$ and $\pi_{\phi} \beta_{\phi}$ have been taken into account in contrast to \eqref{ADM_Hamiltonian}. As a result, we obtain the following expression	
\bqa && \int_{\Sigma_{r}} d^{d} x\ \Bigg\{ \frac{2 \kappa^{2}}{\sqrt{\gamma}} \Bigg(\frac{\delta \mathcal{S}[\gamma_{ij},\phi]}{\delta \gamma_{ij}}\Bigg) \Big(\gamma_{ik} \gamma_{jl} - \frac{1}{d-1} \gamma_{ij} \gamma_{kl}\Big)  \Bigg(\frac{\delta \mathcal{S}[\gamma_{ij},\phi]}{\delta \gamma_{kl}}\Bigg) + \beta_{ij} \frac{\delta \mathcal{S}[\gamma_{ij},\phi]}{\delta \gamma_{ij}} \nn && + \frac{\kappa^{2}}{\sqrt{\gamma}} \Bigg(\frac{\delta \mathcal{S}[\gamma_{ij},\phi]}{\delta \phi}\Bigg)^{2} + \beta_{\phi} \frac{\delta \mathcal{S}[\gamma_{ij},\phi]}{\delta \phi} + \mathcal{V}_{eff}[\gamma_{ij},\phi] \Bigg\} + \frac{\partial \mathcal{S}[\gamma_{ij},\phi]}{\partial r} = 0\ .\label{eq:modifiedHJ} \eqa 
When compared with Eq. (\ref{Hamilton_Jacobi_Eq_FRG}),
\begin{align*}
	&\int d^{d} x\int d^{d} y \Bigg\{ - \frac{\delta \mathcal{S}}{\delta \phi(x,r)} C_{\Lambda}^{Pol.}(x,y,r) \frac{\delta V_{\Lambda}^{Pol.}[\phi]}{\delta \phi(y,r)} + \frac{1}{2} \frac{\delta \mathcal{S}}{\delta \phi(x,r)} C_{\Lambda}^{Pol.}(x,y,r) \frac{\delta \mathcal{S}}{\delta \phi(y,r)} \Bigg\} \notag\\&\hspace*{11cm}+ \frac{\partial \mathcal{S}}{\partial r} = 0\ ,
\end{align*}
the particular term in the generalized Hamilton-Jacobi equation (including the $\beta$-function information) 
\bqa && \int_{\Sigma_{r}} d^{d} x\ \beta_{\phi} \frac{\delta \mathcal{S}[\gamma_{ij},\phi]}{\delta \phi} = \int_{\Sigma_{r}} d^{d} x \frac{1}{\sqrt{\gamma}} \frac{\partial \mathcal{V}_{eff}[\gamma_{ij},\phi]}{\partial \phi} \frac{\delta \mathcal{S}[\gamma_{ij},\phi]}{\delta \phi} \nonumber \eqa 
can be mapped to the term 
\bqa && -\int d^{d} x \int d^{d} y \frac{\delta \mathcal{S}}{\delta \phi(x,r)} C_{\Lambda}^{Pol.}(x,y,r) \frac{\delta V_{\Lambda}^{Pol.}[\phi(x,r)]}{\delta \phi(y,r)}\ , \nonumber \eqa
present in the functional RG flow equation. Accordingly, the Fokker-Planck type functional RG flow equation is given by 		
\bqa && \Big( \frac{\partial }{\partial r} - \int_{\Sigma_{r}} d^{d} x \mathcal{V}_{eff}[\gamma_{ij},\phi] \Big) P_{r}[\gamma_{ij},\phi] \nn &&\hspace*{2cm} = \int_{\Sigma_{r}} d^{d} x \Big\{  \frac{2 \kappa^{2}}{\sqrt{\gamma}} \frac{\delta }{\delta \gamma_{ij}} \Big(\gamma_{ik} \gamma_{jl} - \frac{1}{d-1} \gamma_{ij} \gamma_{kl}\Big) \frac{\delta }{\delta \gamma_{kl}} - \beta_{ij} \frac{\delta }{\delta \gamma_{ij}} \Big\} P_{r}[\gamma_{ij},\phi] \nn &&\hspace*{4cm} + \int_{\Sigma_{r}} d^{d} x \Big\{  \frac{\delta }{\delta \phi} \frac{\kappa^{2}}{\sqrt{\gamma}} \frac{\delta }{\delta \phi} - \beta_{\phi} \frac{\delta }{\delta \phi} \Big\} P_{r}[\gamma_{ij},\phi] . \eqa
On comparing with Eq. (\ref{FRG_Eq}),		
\begin{align*}
	\frac{d }{d \ln \Lambda} P_{\Lambda}[\phi] = \int_{M} d^{d} x \int_{M} d^{d} y \Big\{ C_{\Lambda}^{Pol.}(x,y) \frac{\delta^{2} P_{\Lambda}[\phi]}{\delta \phi(x) \delta \phi(y)} + \frac{\delta}{\delta \phi(x)} \Big( P_{\Lambda}[\phi] C_{\Lambda}^{Pol.}(x,y) \frac{\delta V_{\Lambda}^{Pol.}[\phi]}{\delta \phi(y)} \Big) \Big\}\ ,
\end{align*}
we find that the the drift term $\int_{\Sigma_{r}} d^{d} x \beta_{\phi} \frac{\delta P_{r}[\gamma_{ij},\phi]}{\delta \phi} = \int_{\Sigma_{r}} d^{d} x \frac{1}{\sqrt{\gamma}} \frac{\partial \mathcal{V}_{eff}[\gamma_{ij},\phi]}{\partial \phi} \frac{\delta P_{r}[\gamma_{ij},\phi]}{\delta \phi}$ is consistent with $\int_{M} d^{d} x \int_{M} d^{d} y \frac{\delta}{\delta \phi(x)} \Big( P_{\Lambda}[\phi(x)] C_{\Lambda}^{Pol.}(x,y) \frac{\delta V_{\Lambda}^{Pol.}[\phi(x)]}{\delta \phi(y)} \Big)$.

Finally, we can rewrite the Hamilton-Jacobi equation as a local RG equation as follows	
\bqa && \Bigg\{ \frac{\partial }{\partial r} + \frac{1}{2}\int_{\Sigma_{r}} d^{d} x  (\dot{\gamma}_{ij}+\beta_{ij})\frac{\delta }{\delta \gamma_{ij}} +  \frac{1}{2}\int_{\Sigma_{r}} d^{d} x~ (\dot{\phi}+\beta_{\phi} )\frac{\delta }{\delta \phi} \Bigg\} \mathcal{S} = \int_{\Sigma_{r}} d^{d} x  \mathcal{A}\ . \eqa
The local RG equation can be reorganized as
\bqa && \Bigg\{ \frac{\partial }{\partial r} + \frac{1}{2} \int_{\Sigma_{r}} d^{d} x~ \left(\dot{\gamma}_{ij}\frac{\delta }{\delta \gamma_{ij}} + \dot{\phi}\frac{\delta }{\delta \phi}\right)\Bigg\} \mathcal{S} = \int_{\Sigma_{r}} d^{d} x~ \tilde{\mathcal{A}}\ ,
\eqa
where
\begin{equation}
	\tilde{\mathcal{A}}= \mathcal{A} -\frac{1}{2}  \Big(\beta_{ij}\frac{\delta }{\delta \gamma_{ij}} + \beta_{\phi}\frac{\delta }{\delta \phi}\Big)~\mathcal{S}\ ,
\end{equation}
where the Weyl anomaly is $\mathcal{A} = \mathcal{V}_{eff}[\gamma_{ij},\phi]$. Here, the Hamilton's equations of motion for the `velocity' fields from Eq. \eqref{eq:RGaction} are given by
\bqa && \dot{\gamma}_{ij} - \beta_{ij} = \frac{4 \kappa^{2}}{\sqrt{\gamma}} \Big(\gamma_{ik} \gamma_{jl} - \frac{1}{d-1} \gamma_{ij} \gamma_{kl}\Big) \Big(\frac{\delta \mathcal{S}}{\delta \gamma_{kl}}\Big) , \\ && \dot{\phi} - \beta_{\phi} = \frac{2\kappa^{2}}{\sqrt{\gamma}} \frac{\delta \mathcal{S}}{\delta \phi} , \label{Hamilton_Eq_GR_QFT_Correspondence}\eqa 
respectively. In particular, Eq. (\ref{Hamilton_Eq_GR_QFT_Correspondence}) is consistent with Eq. (\ref{FRG_Gradient_Flow}),
\bqa && \frac{d \phi(x)}{d \ln \Lambda} = \int d^{d} y\ C_{\Lambda}^{Pol.}(x,y) \frac{\delta }{\delta \phi(y)} \Big( \mathcal{S}[\phi] - V_{\Lambda}^{Pol.}[\phi] \Big) , \nonumber \eqa
where Eq. (\ref{Hamilton_Eq_GR_QFT_Correspondence})  can be rewritten as 
\bqa && \dot{\phi} = \frac{2\kappa^{2}}{\sqrt{\gamma}} \frac{\delta}{\delta \phi} \Big( \mathcal{S} + \frac{1}{2\kappa^{2}} \mathcal{V}_{eff}[\gamma_{ij},\phi] \Big)\ . \eqa

Finally, we propose that the relative entropy corresponds to
\bqa && \Sigma = \int_{\Sigma_{R}} d^{d} x \Big( \pi^{ij} \gamma_{ij} + \pi_{\phi} \phi \Big)\ \eqa
in the dual holography framework. It is straightforward to show the monotonicity of this relative entropy functional, using both the Hamilton's equation of motion and the IR boundary condition \cite{RG_Monotonicity_NEQ,RG_Flow_Nonperturbative_String}. We will not discuss this issue further here.

Before closing this section, we give several remarks. The existence of the effective potential and the appearance of the RG $\beta-$function have been discussed on a general ground \cite{SungSik_Holography_I,SungSik_Holography_II}. Furthermore, the gradient RG-flow nature has been speculated in these previous studies. However, the gradient RG-flow nature has not been verified. In appendix, our brute force implementation of the Wilsonian RG transformation confirms the gradient flow nature of the RG $\beta-$function, that is, given by the Luttinger-Ward functional. In Eq. (\ref{HDEFT_Nonperturbative_Local_Approx}), the RG $\beta-$function is given by $\propto \int \frac{d^{d} q}{(2\pi)^{d}} G(p+q,z) D(q,z)$, which can be found by $\propto \frac{\delta}{\delta G(p,z)} \Big( \frac{1}{2} \int \frac{d^{d} q}{(2\pi)^{d}} G(p+q,z) D(q,z) G(p,z) \Big)$.

\section{Discussion and conclusion}

Recently, we derived or more precisely, constructed a dual holography framework based on the one-loop effective potential in a general background \cite{Nonperturbative_Wilson_RG,RG_Monotonicity_NEQ,RG_Flow_Nonperturbative_String}. Such a general background potential originates from the Hubbard-Stratonovich transformation to translate a double-trace interaction term into a single-trace term under an arbitrary background field. This one-loop effective potential in a general background is the only UV information that we need. Then, we obtain the RG flow equation, assuming that the RG $\beta-$function is given by a gradient flow of the effective potential. Resorting to this UV information, we can construct an effective partition function as done in this study, where Gaussian fluctuations for all the coupling functions have been introduced to play the role of noise. It turns out that such noise fluctuations can be derived from irrelevant double-trace deformations \cite{TTbar_Deformation}. As a result, we obtain the dual holography framework in the path integral formulation, where quantum corrections are taken into account in a nonperturbative way.

Here, nonperturbative renormalization effects can be introduced in the following way. First, the one-loop effective potential with a general background field is given in the QFT framework. Then, we obtain the RG $\beta-$function as a gradient flow as discussed before. As a result, the coupling function or the background field is renormalized to RG-flow. This renormalized background field updates the previous one-loop effective potential to RG-flow. This RG step is essential, which does not exist in the perturbative RG procedure. Then, the coupling function is newly updated to renormalize once again. This recursive RG structure serves as the nonperturbative renormalization scheme. We emphasize that this nonperturbative analysis is not exact because we do not perform the path integral for all the dual fields but consider only the saddle-point approximation in the effective bulk partition function.

In this study, we repeat this recursive RG procedure, starting from the functional RG equation instead of following the previous constructive way. In this respect the present study serves as microscopic foundation for our previous microscopic brute-force derivation \cite{Kondo_Holography,RG_Flow_Direct_Calculation} or the recent physics-wise construction \cite{Nonperturbative_Wilson_RG,RG_Monotonicity_NEQ,RG_Flow_Nonperturbative_String} although they turn out to be equivalent. As commented in the previous section, we have to find a superspace formulation to manifest the $\mathcal{N} = 2$ BRST symmetry and obtain the corresponding Ward identities. We will repeat the entropy production calculation \cite{RG_Monotonicity_NEQ,RG_Flow_Nonperturbative_String} in the nonequilibrium thermodynamics perspectives \cite{Entropy_Production} and figure out how this entropy production is consistent with the so called Wess-Zumino consistency condition for the Weyl anomaly in the local RG equation \cite{Local_RG_I,Local_RG_II,Local_RG_III}, also being responsible for the monotonicity of the RG flow.

\vspace*{10mm}

{\footnotesize \noindent {\bf Acknowledgements:} K.-S. K. was supported by the Ministry of Education, Science, and Technology (Grant No. RS-2024-00337134) of the National Research Foundation of Korea (NRF).\footnotesize }

\appendix
\section{Explicit derivation of the path integral formulation for the functional RG analysis}\label{appendix1}

In this appendix, we present the explicit derivation of the path integral formulation for the functional RG analysis. We start from an $O(N)$ vector model whose partition function is
\bqa && Z = \int D \varphi_{a}(x) \exp\Big[ - \int d^{d} x \Big\{ \sum_{a = 1}^{N} \Big( \partial_{\mu} \varphi_{a}(x) \Big)^{2} + m^{2} \sum_{a = 1}^{N} \varphi_{a}^{2}(x) \nn && \hspace*{8cm}+ \frac{u}{2 N}  \left(\sum_{a= 1}^{N}\varphi_{a}^{2}(x) \right)^2 \Big\} \Big] . \eqa
Here, $a$ denotes flavor indexes for the $O(N)$ symmetry running from $1$ to $N$.

To study the dynamical propagators, we introduce a collective bilocal field as
\bqa && G(x,x') = \frac{1}{N} \sum_{a = 1}^{N} \varphi_{a}(x) \varphi_{a}(x') \equiv N^{-1} \varphi_{a}(x) \varphi_{a}(x')\ , \eqa
where the Einstein summation convention for the spin summation has been used in the last equality. Then, we can rewrite the above expression as follows
\bqa && Z = \int D \varphi_{a}(x) D G(x,x') D \Sigma(x',x) \nn && \exp\Big[ - \int d^{d} x \Big\{ \Big( \partial_{\mu} \varphi_{a}(x) \Big)^{2} + m^{2} \varphi_{a}^{2}(x) + \frac{N u}{2} G^{2}(x,x) \Big\} \nn && - i \int d^{d} x \int d^{d} x' \Sigma(x',x) \Big( N G(x,x') - \varphi_{a}(x) \varphi_{a}(x') \Big) \Big] .\label{eq:GF} \eqa
Here, $\Sigma(x',x)$ is a Lagrange multiplier field to impose the Green's function definition above. Note that in \eqref{eq:GF}, only the final line contains all bilocal terms.

One can perform the Gaussian integral for the Green's function field $G(x,x')$ splitting it into  diagonal and off-diagonal parts. The off-diagonal Gaussian integrals over $G(x,x')$ produce delta functions up to normalization factors and assuming a flat functional measure. It forces $\Sigma(x',x)$ to vanish for all $x \neq x'$, i.e.\ $\Sigma$ has support only on the diagonal $x = x'$. The most general distribution with this property is $\Sigma(x',x) = \sigma(x)\,\delta^d(x-x')$, where $\sigma(x)$ is a local field encoding the strength of $\Sigma$ along the diagonal. Therefore, we obtain
\begin{align}
	Z = \int D \varphi_{a}(x) D \Sigma(x) \exp\Big[ - \int d^{d} x \Big\{ \varphi_{a}(x) \Big( - \partial_{\mu}^{2} + m^{2}- i \Sigma(x) \Big) \varphi_{a}(x) + \frac{N}{2 u} \Sigma^{2}(x) \Big\} \Big] .\label{eq:GFs}    
\end{align}
This is our starting point for the RG transformation. Note that, locality of the field $\Sigma(x)$ in \eqref{eq:GFs} renders the quadratic operator non-diagonal in momentum space,
$$\int\frac{d^dp}{(2\pi)^d}\frac{d^dp'}{(2\pi)^d}\tilde{\varphi}_{a}(p)\Sigma(-p-p')\tilde{\varphi}_{a}(p')\ ,$$
so a standard momentum-shell decomposition is not directly convenient. To overcome this, we will introduce a decoupled auxiliary field $\psi_a(x)$ following \cite{SungSik_Holography_III,SungSik_Holography_IV}, 
\bqa && Z = \int D \psi_{a}(x) D \varphi_{a}(x) D \Sigma(x) \nn && \exp\Big[-\int d^{d} x \Big\{ \varphi_{a}(x) \Big( - \partial_{\mu}^{2} + m^{2}- i \Sigma(x) \Big) \varphi_{a}(x) + M^{2} \psi_{a}^{2}(x) + \frac{N}{2 u} \Sigma^{2}(x) \Big\} \Big] . \eqa
This enlarges the field space. 
By appropriately choosing the transformation coefficients, one can separate the fields into components associated with low- and high-momentum modes. This construction restores a well-defined coarse-graining procedure and enables the implementation of the Wilsonian RG in a manner consistent with the emergent holographic description.

First we separate both the fields in high- and low-energy modes by performing a linear transformation
\bqa && \varphi_{a}(x) \longrightarrow \phi_{a}(x) + \Phi_{a}(x) , ~~~~~ \psi_{a}(x) \longrightarrow c_{\phi} \phi_{a}(x) + c_{\Phi} \Phi_{a}(x) . \eqa
The field $\phi_a$ represents low-energy, slow-moving modes while $\Phi_a$ encodes the high-energy, fast moving modes which are to be integrated out. The fine-tuned coefficients $c_{\phi}$ and $c_{\Phi}$ ensure a well-defined momentum space decoupling of high and low-energy modes which can subsequently be integrated out in a well-defined manner. This essentially reproduces the essential structure of the Wilsonian RG entirely in real space.

Here, both coefficients of $c_{\phi}$ and $c_{\Phi}$ are given by
\bqa 
\label{eq:cmuexpr}
c_{\phi} = \frac{\cal{O}}{\mu M} , ~~~~~ c_{\Phi} = - \frac{\mu}{M} ,~~~~~ \mu = \sqrt{\frac{\cal{O}}{e^{2 \alpha d z} - 1} }, \eqa
with $${\cal{O}}= - \partial_{\mu}^{2} + m^{2}- i \Sigma(x)$$
Here the coefficients are operator valued. These are determined by the fact that there are no crossing terms between high-energy fields and low-lying modes as follows
\begin{equation} \phi_{a}(x)~ {\cal{O}}~ \phi_{a}(x) + M^{2} \psi_{a}^{2}(x) = e^{2 \alpha d z} \phi_{a}(x)~{\cal{O}} ~\phi_{a}(x) + \frac{e^{2 \alpha d z}}{e^{2 \alpha d z} - 1} \Phi_{a}(x) ~{\cal{O}}~ \Phi_{a}(x)\ . \end{equation}
Here, $\alpha$ has been introduced as a control parameter for the RG transformation. 
As a result, we obtain,
\bqa && Z = \int D \Phi_{a}(x) D \phi_{a}(x) D \Sigma(x) ~ \mathcal{J} \nn && \exp\Big[ - \int d^{d} x \Big\{ \phi_{a}(x) {\cal{O}} \phi_{a}(x)  + \frac{1}{e^{2 \alpha d z} - 1} \Phi_{a}(x) {\cal{O}} \Phi_{a}(x) + \frac{N}{2 u} \Sigma^{2}(x) \Big\} \Big] . \label{eq:PF1} \eqa
where, we performed the following rescaling 
\bqa && \phi_{a}(x) \longrightarrow e^{- \alpha d z} \phi_{a}(x) , ~~~~~ \Phi_{a}(x) \longrightarrow e^{- \alpha d z} \Phi_{a}(x) \label{eq:fieldrescale}\eqa 
with $\partial_{\mu} \alpha = 0$, to ensure that the quadratic operator retains its canonical form, while the fast field acquires the large coefficient $1/(e^{2\alpha dz}-1)\sim 1/(2\alpha dz)$. This condition guaranties that the effective action remains within the same functional class, allowing the RG flow to be interpreted as an evolution of the coupling parameters. The resulting partition function in \eqref{eq:PF1} demonstrates a Wilsonian structure containing slow fields with canonical normalization and fast fields suppressed. 
For the rescaling in \eqref{eq:fieldrescale}, using a regulated trace (e.g. heat-kernel regularization),
one finds that the measure generates local contributions to the effective action. These terms are polynomial in $\Sigma$ and correspond to standard RG counterterms.
Now, it is straightforward to perform the Gaussian integral for high-energy fields.

As a result, we obtain
\bqa && Z = \int D \phi_{a}(x) D \Sigma(x) \nn && \hspace*{1.7cm}\exp\left[ - \int d^{d} x \Big\{ \phi_{a}(x) {\cal{O}} \phi_{a}(x) + \frac{N}{2 u} \Sigma^{2}(x)  \Big\}- \frac{N}{2} \alpha d z \text{Tr}\ln ({\cal{O}}) \right]\ . \eqa
Next, we perform an RG transformation for the collective field $\Sigma(x)$ in a fashion similar to that of $\varphi_a$ as outlined earlier. First, we introduce an \textit{auxiliary field} $\chi(x)$ which affects the theory trivially\footnote{Such a trivial quadratic piece of an auxiliary field $\chi(x)$ affects the partition function only to an overall normalization thus keeping the theory intact as before.}

\begin{equation}
	\begin{aligned}
		Z&=\int D \phi_{a}(x) D\Sigma(x) D \chi(x)\\
		&\hspace*{1.5cm}\exp\left[-\int d^{d}x \Big\{ \phi_{a}(x) {\cal{O}} \phi_{a}(x) + \frac{N}{2 u} \Sigma^{2}(x) + \frac{N}{2 \tilde{u}} \chi^{2}(x) \Big\} -\frac{N}{2} \alpha d z \text{Tr}\ln {\cal{O}}\right]\ .
	\end{aligned}
\end{equation}
In the same way as the RG transformation for the matter field, we separate fast degrees of freedom (denoted by $\tilde{\Sigma}^{(1)}$) from slow ones (denoted by $\Sigma^{(0)}$) by the following field redefinition
\bqa && \Sigma(x) \longrightarrow \Sigma^{(0)}(x) + \tilde{\Sigma}^{(1)}(x)\ \ , ~~~ \chi(x) \longrightarrow c_{\Sigma} \Sigma^{(0)}(x) + c_{\tilde{\Sigma}} \tilde{\Sigma}^{(1)}(x)\ . \eqa
Similar to before, we again demand decoupling between the sectors $\Sigma^{(0)}$ and $\tilde{\Sigma}^{(1)}$. Plugging the above into the action, the relevant part becomes:
\begin{equation}
	\begin{aligned}
		&\frac{N}{2 u} \Sigma^{2}(x) + \frac{N}{2 \tilde{u}} \chi^{2}(x) =\\& \frac{N}{2}\left(\frac{1}{u}+\frac{c_{\Sigma}^2}{\tilde{u}}\right)(\Sigma^{(0)})^{ 2}+\frac{N}{2}\left(\frac{1}{u}+\frac{c_{\tilde{\Sigma}}^2}{\tilde{u}}\right)(\tilde{\Sigma}^{(1)})^{ 2}+N\left( \frac{1}{u}+\frac{c_{\Sigma}c_{\tilde{\Sigma}}}{\tilde{u}}\right)\Sigma^{(0)}\tilde{\Sigma}^{(1)}
	\end{aligned}
\end{equation}
The quadratic piece in the slow moving mode must be proportional to the original one. Denoting the RG flow parameter by $\beta$ and and demanding complete decoupling of fast and slow moving modes leads us to two equations:
\begin{equation}
	\frac{1}{u}+\frac{c_{\Sigma}c_{\tilde{\Sigma}}}{\tilde{u}} =0 \quad \text{and} \quad \frac{N}{2}\left(\frac{1}{u}+\frac{c_{\Sigma}^2}{\tilde{u}}\right)=e^{2\beta dz}\frac{N}{2u}\ .
\end{equation}
Solving for the coefficients $c_{\Sigma}$ and $c_{\tilde{\Sigma}}$ gives
\bqa && c_{\Sigma} = \frac{\sqrt{\tilde{u}}}{\mu u} , ~~~~~ c_{\tilde{\Sigma}} = - \mu \sqrt{\tilde{u}} , ~~~~~ \mu = \frac{1}{\sqrt{u} \sqrt{e^{2 \beta d z} - 1}} , \eqa
Note that here the Jacobian factor due to the transformation is \textit{independent of fields} and hence \textit{trivial}. Thus, we do not need to include further in our computations. We thus obtain the effective theory to be given by
\begin{equation}
	\begin{aligned}
		Z&=\int D \phi_{a}(x) D \tilde{\Sigma}^{(1)}(x) D \Sigma^{(0)}(x)\\
		&\hspace*{1cm}\exp\left[-\int d^{d}x \Big\{ \phi_{a}(x) {\cal{O}} \phi_{a}(x)+ e^{2\beta dz}\frac{N}{2 u}( \Sigma^{(0)})^{2}(x) \right.\\
		&\hspace*{4cm}\left.+ \frac{N}{2u}\frac{e^{2\beta dz}}{e^{2\beta dz}-1} (\tilde{\Sigma}^{(1)})^{2}(x)  \Big\} -\frac{N}{2} \alpha d z \text{Tr}\ln \mathcal{O}\right]\ ,
	\end{aligned}
\end{equation}
where the operator $\mathcal{O}$ should be read as
\begin{equation}
	\mathcal{O} \equiv -\partial_\mu^2 + m^2 - i\Sigma^{(0)}(x) -i\tilde{\Sigma}^{(1)}\ .
\end{equation}
Furthermore, we now perform a rescaling of the field $\Sigma^{(0)}$ and $\tilde{\Sigma}^{(1)}$ as
\begin{equation}
	\Sigma^{(0)}(x) \longrightarrow e^{- \beta d z} \Sigma^{(0)}(x)\ \quad \text{and} \quad\  \tilde{\Sigma}^{(1)}(x) \longrightarrow e^{- \beta d z} \tilde{\Sigma}^{(1)}(x)\ ,
\end{equation}
we end up with
\begin{equation}
	\begin{aligned}
		Z&=\int D \phi_{a}(x) D \tilde{\Sigma}^{(1)}(x) D \Sigma^{(0)}(x)\\
		&\hspace*{1cm}\exp\left[-\int d^{d}x \Big\{ \phi_{a}(x) {\cal{O}}' \phi_{a}(x)+ \frac{N}{2 u} (\Sigma^{(0)})^{2}(x)+ \frac{N}{2u}\frac{1}{e^{2\beta dz}-1} (\tilde{\Sigma}^{(1)})^{2}(x)  \Big\}\right.\\
		&\hspace*{10cm}\left.-\frac{N}{2} \alpha dz \text{Tr}\ln \mathcal{O}'\right]\ ,
	\end{aligned}
\end{equation}
where now, the operator $\mathcal{O}'$ is
\begin{equation}
	\mathcal{O}' \equiv -\partial_\mu^2 + m^2 - ie^{-\beta dz}\Sigma^{(0)}(x) -ie^{-\beta dz}\tilde{\Sigma}^{(1)}
\end{equation}
In order to now integrate over $\tilde{\Sigma}^{(1)}$, we need to first expand out the above expression in small $dz$. We can first expand the $\text{Tr} \ln$ term as
\begin{equation}
	\begin{aligned}
		\text{Tr} \ln \mathcal{O}'&=  \text{Tr} \ln (\mathcal{O}_0+V)\\&= \text{Tr} \ln \mathcal{O}_0+\text{Tr}(G_0 V)-\frac{1}{2} \text{Tr}(G_0 VG_0 V)+\dots
	\end{aligned}
\end{equation}
where 
$\mathcal{O}_0=-\partial_\mu^2 + m^2 - i\Sigma^{(0)}(x),\, G_0=\mathcal{O}_0^{-1},\,V=-ie^{-\beta dz}\tilde{\Sigma}^{(1)}$. 
Considering the small $dz$ expansion such that $dz \to 0$, $N \to \infty$ but $N dz$ is $O(1)$, we keep terms upto \textit{linear order} in $dz$ thus obtaining,
\begin{equation}
	\begin{aligned}
		\frac{N}{2}\alpha dz  \text{Tr} \ln \mathcal{O}' &=\frac{\alpha}{2}N dz \text{Tr}\ln \mathcal{O}_0-\frac{i\alpha}{2} N dz \text{Tr}(G_0 \tilde{\Sigma}^{(1)})+\frac{\alpha}{4}Ndz \text{Tr}(G_0\tilde{\Sigma}^{(1)}G_0\tilde{\Sigma}^{(1)})+ \cdots\ .    
	\end{aligned}
\end{equation}
The traces in the above equation are explicitly given by
\begin{equation}
	\text{Tr}\big(G_0\tilde{\Sigma}^{(1)}\big) = \int d^d x\, d^d x'\; \mathcal{O}_0^{-1}(x, x')\,\tilde{\Sigma}^{(1)}(x')\,\delta^{(d)}(x' - x)\,.
\end{equation}
and
\begin{equation}
	\text{Tr}\Big(G_0\tilde{\Sigma}^{(1)}G_0\tilde{\Sigma}^{(1)}\Big) = \int d^d x d^d y\ \tilde{\Sigma}^{(1)}(y)\mathcal{O}_0^{-1}(x, y)\mathcal{O}_0^{-1}(y,x)\tilde{\Sigma}^{(1)}(x)\ .\footnote{Once the trace is converted into explicit integrals, the individual elements appearing in the integrand are c-numbers and can be reordered in any way.}
\end{equation}
It must be noted that the inverse of the local differential operator $\mathcal{O}_0(x)$ i.e. $\mathcal{O}_0^{-1}(x) \equiv G_0(x,y)$ is infact a \textit{non-local} Green's function satisfying
\begin{equation}
	\left[-\partial_{\mu}^2+m^2-i\Sigma^{(0)}(x) \right]G_0(x,y)=\delta^{(d)}(x-y)\ .
\end{equation}
The source is localized at $y$ while the response propagates to $x$ through the differential operator.

Performing the expansion, we eventually end up with the effective theory explicitly given by 
\begin{equation}
	\begin{aligned}
		Z&=\int D \phi_{a}(x) D \tilde{\Sigma}^{(1)}(x) D \Sigma^{(0)}(x)\\
		&\exp \left[-\int d^dx\ \left\{\phi_a(x)\left(-\partial_{\mu}^2+m^2-i\Sigma^{(0)}(x)-i\tilde{\Sigma}^{(1)}(x) \right)\phi_a(x)\right.\right.\\
		&\hspace*{7.5cm}\left.+\frac{N}{2u}(\Sigma^{(0)}(x))^2+\frac{N}{4u\beta dz}(\tilde{\Sigma}^{(1)}(x))^2\right\}\\
		&-\int d^dx d^dy \left\{\frac{N}{4}\alpha dz\ \tilde{\Sigma}^{(1)}(y)G_0(x,y)G_0(y,x)\tilde{\Sigma}^{(1)}(x)-i\frac{ N}{2}\alpha dz G_0(x,y)\tilde{\Sigma}^{(1)}(y)\delta^{(d)}(x-y)\right\}\\
		&\hspace*{7cm}\left.-\frac{ N}{2}\alpha dz\int d^dx \ln(-\partial_{\mu}^2+m^2-i\Sigma^{(0)}(x))\right]\ .
	\end{aligned}
\end{equation}

In the above we have kept \textit{the most dominant terms} in the first two lines while in the third line, we have kept terms upto linear order in $dz$. Before we perform the integration of $\tilde{\Sigma}^{(1)}(x)$ field, we perform a shift $\tilde{\Sigma}^{(1)}(x) \longrightarrow \tilde{\Sigma}^{(1)}(x) - \Sigma^{(0)}(x)$ and drop the `tilde' for the sake of brevity to obtain
\begin{equation}
	\begin{aligned}
		Z&=\int D \phi_{a}(x) D \Sigma^{(0)}(x) D \Sigma^{(1)}(x)\\
		&\exp \left[-\int d^dx \left\{\phi_a(x)\left(-\partial_{\mu}^2+m^2-i\Sigma^{(1)}(x)\right)\phi_a(x)+\frac{N}{2u}(\Sigma^{(0)}(x))^2\right\}\right.\\
		&+\int d^dx\ d^dy\ i\frac{\alpha Ndz}{2}G_0(x,y)\left({\Sigma}^{(1)}(y)-\Sigma^{(0)}(y)\right)\delta^{(d)}(x-y)\\
		&-\int d^dx d^dy\ \frac{N}{4\beta dz}\left(\Sigma^{(1)}(y)-\Sigma^{(0)}(y)\right)\mathcal{K}_u(x,y)\left(\Sigma^{(1)}(x)-\Sigma^{(0)}(x)\right)\\
		&\hspace*{5cm}\left.-\frac{\alpha N dz}{2}\int d^dx \ln(-\partial_{\mu}^2+m^2-i\Sigma^{(0)}(x))\right]\ .
	\end{aligned}
\end{equation}
where
\begin{equation}
	\mathcal{K}_u(x,y)=\mathcal{K}_u(y,x)=\frac{1}{u}+\alpha \beta (dz)^2G_0(x,y)G_0(y,x)\ . 
\end{equation}
Diagramatically speaking, the quadratic term in $G_0$ actually represents the \textit{polarization bubble}. As seen from the \textit{leading order} piece in $\phi_{\alpha}(x) \Big( - \partial_{\mu}^{2} + m^{2} - i \Sigma^{(1)}(x) \Big) \phi_{\alpha}(x)$, $\Sigma^{(0)}(x)$ flows into $\Sigma^{(1)}(x)$ after an infinitesimal RG step.

To obtain the corresponding continuum expression above, we multiply by appropriate factors of $dz$ to obtain
\begin{equation}
	\begin{aligned}
		Z&=\int D \phi_{a}(x) D \Sigma^{(0)}(x) D \Sigma^{(1)}(x)\exp \left[-\int d^dx \left\{\phi_a(x)\left(-\partial_{\mu}^2+m^2-i\Sigma^{(1)}(x)\right)\phi_a(x)\right.\right.\\
		&\hspace*{2cm}\left. +\frac{N}{2u}(\Sigma^{(0)}(x))^2-\frac{\alpha N dz}{2} \ln (-\partial_{\mu}^2+m^2 -i \Sigma^{(0)}(x))\right\}\\
		&+\int d^dx\ d^dy\ i\frac{\alpha N}{2}(dz)^2G_0(x,y)\frac{\Sigma^{(1)}(y)-\Sigma^{(0)}(y)}{dz}\delta^{(d)}(x-y)\\
		&\left.-\int d^dx\ d^dy\ \frac{N}{4\beta}dz\frac{\left(\Sigma^{(1)}(y)-\Sigma^{(0)}(y)\right)}{dz}\mathcal{K}_u(x,y)\frac{\left(\Sigma^{(1)}(x)-\Sigma^{(0)}(x)\right)}{dz}\right]\ .
	\end{aligned}
\end{equation}
Furthermore, repeating these RG transformations and replacing $dz$ by $\varepsilon$, we get an effective dual theory,
\begin{equation}
	\begin{aligned}
		Z&=\int D\phi_{a}(x) D\Sigma(x,z)
		\exp \left[-\int d^dx \left\{ \phi_a(x)\left(-\partial_{\mu}^2+m^2-i\Sigma(x,z_f)\right)\phi_a(x)\right.\right.\\&\left.\left.+\frac{N}{2u}(\Sigma(x,0))^2\right\}
		-\frac{\alpha N}{2}\int d^dx\ dz\ \ln(-\partial_{\mu}^2+m^2-i\Sigma(x,z))\right. \\
		&+i\frac{\alpha N}{2}\varepsilon\int d^dx\ d^dy\ dz\ G(x,y;z)\partial_z \Sigma(y,z)\delta^{(d)}(x-y)\\
		&\left.-\frac{N}{4\beta}\int d^dx\ d^dy\ dz\ \partial_z \Sigma(y,z)\mathcal{K}_u(x,y;z)\partial_z \Sigma(x,z)\right]\ .
	\end{aligned}
\end{equation}
The above can alternatively written as:
\begin{equation}
	\begin{aligned}
		Z&=\int D\phi_{a}(x) D\Sigma(x,z)\\
		&\exp \left[-\int d^dx \left\{\phi_a(x)\left(-\partial_{\mu}^2+m^2-i\Sigma(x,z_f)\right)\phi_a(x)+\frac{N}{2u}(\Sigma(x,0))^2\right\}\right.\\
		&\hspace*{6cm}-\frac{\alpha N}{2}\int d^dx\ dz\ \ln(-\partial_{\mu}^2+m^2-i\Sigma(x,z)) \\
		& -\frac{N}{4\beta}\int d^dx\ d^dy\ dz\ \left\{\partial_z \Sigma(y,z)-i \alpha \beta \varepsilon G(y,x;z)\mathcal{K}_u^{-1}(x,y;z) \right\}\mathcal{K}_u(x,y;z)\\
		&\hspace*{6cm}\times \left\{\partial_z \Sigma(x,z)-i \alpha \beta \varepsilon G(x,y;z)\mathcal{K}_u^{-1}(x,y;z)\right\}\\
		&\hspace*{4cm}\left.-\varepsilon^2 \frac{N \alpha^2 \beta}{4}\int d^dx\ d^dy\ dz\ G(y,x;z)\mathcal{K}_u^{-1}(x,y;z)G(x,y;z)\right]\ .
	\end{aligned}\label{k}
\end{equation}
Note that in writing the above, we have used the symmetry of the function $\mathcal{K}_u(x,y;z)=\mathcal{K}_u(y,x;z)$. In Eq. \eqref{k}, the first parentheses in exponential represent the boundary contributions evaluated at the IR boundary $z=z_f$ and the UV boundary $z=0$. The subsequent terms can be interpreted as the bulk contribution in the emergent bulk dual.

Therefore, based on the above microscopic derivation, we propose a general structure for the \textit{holographic dual effective field theory} as follows 
\begin{equation}
	\begin{aligned}
		Z&=\int D\phi_a(x)D\Sigma(x,z)\ \exp \left[-\int d^dx \left\{ \int d^dy\ \phi_a(x)G^{-1}(x,y,z_f)\phi_a(y)+\frac{N}{2u}(\Sigma(x,0))^2 \right\}\right.\\
		&-\frac{N}{4\beta}\int d^dx\ d^dy\ \int_0^{z_f}dz\ \Big(\partial_z\Sigma(y,z)-i\alpha \beta \varepsilon G(y,x;z)D(x,y;z)\Big)D^{-1}(x,y;z)\\
		&\hspace*{6cm}\times \Big(\partial_z \Sigma(x,z)-i\alpha \beta \varepsilon G(x,y;z)D(x,y;z)\Big)\\
		&\left.-\frac{N\alpha}{2}\int d^dx\ d^dy  \ \int_0^{z_f}dz\ \ln G^{-1}(x,y;z)\right.\\&\left.-\varepsilon^2 \frac{N\alpha^2 \beta}{4}\int d^dx\ d^dy\ \int_0^{z_f}dz\ G(x,y;z)D(x,y;z)G(y,x;z)\right]\ ,
	\end{aligned}
\end{equation}
where the Green's functions are given by
\begin{eqnarray}
	\Big( -\partial_{\mu}^2+m^2-i\Sigma(x,z)\Big)G(x,y;z)&=&\delta^{(d)}(x-y)\ ,\\
	\Big(\frac{1}{u}+\alpha \beta \varepsilon^2 G(x,y;z)G(y,x;z) \Big)D(x,y;z)&=&\delta^{(d)}(x-y)\ .
\end{eqnarray}
Taking the $\varepsilon \rightarrow 0$ limit and performing the Fourier transformation, we obtain
\begin{equation}
	\label{HDEFT_Nonperturbative_Local_Approx}
	\begin{aligned}
		Z&=\int D\phi_a(p)D\Sigma(p;z) \\ & \exp \left[-\int d^dp\ \int d^dp'\ \phi_a(p)G^{-1}(p^{2};\Sigma(p-p',z_{f}))\phi_a(-p') - \frac{N}{2u}\int d^dp\ \Sigma(p,0)\Sigma(-p,0)\right.\\
		&-\frac{N}{4 \beta u}\int d^dp \ \int_0^{z_f}dz\ \Big(\partial_z \Sigma(p,z)\Big) \Big(\partial_z \Sigma(-p;z)\Big)\\
		&\left.- \frac{N\alpha}{2} \int d^dp\ \int d^dp'\ \int_0^{z_f}dz\ \ln G^{-1}(p^{2};\Sigma(p-p',z)) \right]\ ,
	\end{aligned}
\end{equation}
where the Green's functions are given by
\bqa && \Big( p^{2} + m^{2} - i \Sigma(p-p',z) \Big) G(p^{2};\Sigma(p-p',z)) = 1 . 
\eqa
This reproduces the holographic dual effective field theory, Eq. (\ref{HDEFT_Nonperturbative}).

\end{document}